\documentclass[preprint,a4paper,notoc]{JHEP3}

\usepackage{amsmath}
\usepackage{epsfig}
\usepackage{multicol}
\usepackage{psfrag}
\usepackage{amssymb}
\usepackage{bbm}
\usepackage{cite}
\usepackage{subfigure}

\newcommand{\parbreak}{\vskip 0.12cm}
\newcommand{\Eq}[1]{{Eq.~(\ref{#1})}}

\newcommand{\id}{1\hspace{-0,9ex}1}
\newcommand{\Oa}{{${\rm O}(a)\ $}}

\newcommand{\Oasquare}{{${\rm O}(a^2)\ $}}
\newcommand{\ZAstat}{{Z^{\rm stat}_{\rm\scriptscriptstyle A}}}
\newcommand{\ZVstat}{{Z^{\rm stat}_{\rm\scriptscriptstyle V}}}
\newcommand{\cAstat}{{c^{\rm stat}_{\rm\scriptscriptstyle A}}}
\newcommand{\bAstat}{{b^{\rm stat}_{\rm\scriptscriptstyle A}}}
\newcommand{\cVstat}{{c^{\rm stat}_{\rm\scriptscriptstyle V}}}
\newcommand{\bVstat}{{b^{\rm stat}_{\rm\scriptscriptstyle V}}}
\newcommand{\cAstatone}{{c^{\rm stat(1)}_{\rm\scriptscriptstyle A}}}
\newcommand{\bAstatzero}{{b^{\rm stat(0)}_{\rm\scriptscriptstyle A}}}
\newcommand{\bAstatone}{{b^{\rm stat(1)}_{\rm\scriptscriptstyle A}}}
\newcommand{\cVstatone}{{c^{\rm stat(1)}_{\rm\scriptscriptstyle V}}}
\newcommand{\bVstatone}{{b^{\rm stat(1)}_{\rm\scriptscriptstyle V}}}
\newcommand{\lp}{{\scriptscriptstyle +}}
\newcommand{\lm}{{\scriptscriptstyle -}}
\newcommand{\cP}{{\cal P}}

\newcommand{\re}{{\rm e}}
\newcommand{\bp}{{\mathbf{p}}}
\newcommand{\bu}{{\mathbf{u}}}
\newcommand{\bv}{{\mathbf{v}}}
\newcommand{\by}{{\mathbf{y}}}
\newcommand{\bz}{{\mathbf{z}}}
\newcommand{\bx}{{\mathbf{x}}}
\newcommand{\heavy}{\psi_{\rm \phantom{\bar{h}^\dagger}\hspace{-0.28cm}h}}
\newcommand{\heavyb}{{\overline{\psi}}_{\rm h}}
\newcommand{\aheavy}{\psi_{{\rm \phantom{\bar{h}^\dagger}\hspace{-0.28cm}\bar{h}}}}
\newcommand{\aheavyb}{\overline{\psi}_{\bar{\rm h}}}

\title{${\rm O}(a)$ improvement of the HYP static axial and vector currents 
at one-loop order of perturbation theory}

\author{}
\author{\epsfig{figure=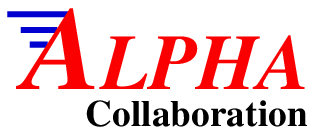,width=2.5cm}}
\author{}

\author{
  Alois Grimbach\\
  Fachbereich C, Bergische Universit\"at Wuppertal, 42097 Wuppertal, Germany\\
  E-mail: \email{grimbach@physik.uni-wuppertal.de}}

\author{
  Damiano Guazzini\\
  DESY, Theory Group, Platanenallee 6, D-15738 Zeuthen, Germany\\
  E-mail: \email{damiano.guazzini@desy.de}}

\author{
  Francesco Knechtli\\
  Fachbereich C, Bergische Universit\"at Wuppertal, 42097 Wuppertal, Germany\\
  E-mail: \email{knechtli@physik.uni-wuppertal.de}}

\author{
  Filippo Palombi\\
  CERN, Physics Department, TH Division, CH-1211 Geneva, Switzerland\\  
  E-mail: \email{filippo.palombi@cern.ch}}

\preprint{CERN-PH-TH/2008-010 \\
{SFB/CPP-08-09}\\
{WUB/08-01}\\[2.0ex]
{February 2008}}

\abstract{
  We calculate analytically the improvement coefficients of the static axial and vector 
  currents in ${\rm O}(a)$ improved lattice QCD at one-loop order of perturbation 
  theory. The static quark is described by the hypercubic action, previously introduced in 
  the literature in order to improve the signal-to-noise ratio of static observables. 
  Within a Schr\"odinger Functional setup, we derive the Feynman rules of the hypercubic 
  link in time-momentum representation. The improvement coefficients are obtained from 
  on-shell correlators of the static axial and vector currents. As a by-product, we localise 
  the minimum of the static self-energy as a function of the smearing parameters of the 
  action at one-loop order and show that the perturbative minimum is close to its 
  non-perturbative counterpart. 
}


\begin{document}

\section{Introduction}
The hypercubic (HYP) link has been introduced a few years ago 
in order to study the static quark/anti-quark potential with 
improved statistical precision \cite{Hasenfratz:2001hp}. In this 
context it proved to be remarkably effective at large quark 
distances, where a description of the static propagators in terms 
of thin gauge links was known to lead to an excessive noise. 
The origin of the statistical improvement has been subsequently 
identified as due to a strong reduction of the static self-energy 
for appropriate choices of the HYP smearing parameters 
\cite{DellaMorte:2005yc}. 

\vspace{0.1cm}

Since the binding energy of static mesons is amplified 
by self-energy contributions, the adoption of the HYP 
link as a parallel transporter in the static action, originally 
introduced in ref.~\cite{Eichten:1989zv}, has allowed for lattice 
simulations with smaller binding potentials, thus triggering an 
exponential improvement of the signal-to-noise ratio of meson 
correlators at large time distances. This helped significantly in 
several applications of Heavy Quark Effective Theory (HQET) 
\cite{DellaMorte:2006cb,Guazzini:2007bu,DellaMorte:2007ij,Palombi:2007dr,Palombi:2007yh,%
Dimopoulos:2007ht}. 

\vspace{0.1cm}

On a theoretical side, the HYP smearing procedure looks superior 
to other techniques in that it mixes gauge links within hypercubes 
attached to the original link only, which allows to preserve 
locality to a high degree. 

\vspace{0.1cm}

In spite of the statistical gain, the HYP link brings on an increase 
of operational complexity, which in some applications has compelled 
to resort to approximations. For instance, this is the case 
with the \Oa improvement of the static-light bilinear operators, 
which has been addressed to in \cite{DellaMorte:2005yc,Palombi:2007dt}. 
In these papers the improvement of the static axial and vector 
currents is obtained by imposing that on-shell renormalised 
correlators of the currents with different choices of the HYP smearing 
parameters differ by \Oasquare effects. Obviously, this procedure 
allows to relate the operator improvement coefficients corresponding 
to different static discretisations, but in order to isolate all of 
them, the improvement coefficients must be known for at least one 
choice of the smearing parameters. 

\vspace{0.1cm}

An approximate solution has been obtained by using a one-loop perturbative 
evaluation of the improvement coefficients with static discretisations 
characterised by simpler lattice Feynman rules, such as the original 
Eichten-Hill (EH) one or the so-called APE discretisation 
(see \cite{DellaMorte:2005yc} and sect.~2 for definitions). The approximate 
solutions thus obtained for other choices of the HYP parameters are 
therefore affected by ${\rm O}(g_0^4)$ systematic errors. As such,  
they have to be regarded as effective estimates, which are neither purely
one-loop perturbative, nor fully non-perturbative. Of course, the amount 
of the ${\rm O}(g_0^4)$ terms can be assessed  only if the one-loop value 
of such improvement coefficients is known exactly.

\vspace{0.1cm}

The aim of the present paper is precisely the analytical determination
of the improvement coefficients of the static axial and vector currents
at one-loop order of perturbation theory (PT) for the most convenient choices 
of the HYP link. In order to accomplish this, we adopt a formalism based 
on the Schr\"odinger Functional (SF) \cite{Luscher:1992an,Luscher:1996vw}, 
derive the lattice Feynman rules of the HYP link at one-loop order of 
PT in time-momentum representation, and follow 
\cite{Palombi:2007dt,Kurth:2000ki} as for the choice of the operator 
improvement conditions. 

\vspace{0.1cm}

As a by-product, we also calculate the static self-energy at one-loop order 
as a function of the smearing parameters and perform a numerical minimisation 
of it. Remarkably, the minimum of the self-energy is found for a choice of 
the smearing parameters which is very close to what is nowadays called the 
HYP2 action \cite{DellaMorte:2005yc}, thus showing that the self-energy is 
dominated by perturbation theory%
\footnote{The perturbative minimisation has been performed originally by 
R.~Hoffmann using periodic boundary conditions, but it has never been 
published. We did it independently and are grateful to him for private 
communications.}.  

\vspace{0.1cm}

The paper is organised as follows. In sect.~2 we review some basic definitions 
and introduce our notation for the HYP static action. Sect.~3 is devoted to a 
discussion of the effects of the SU(3) projection, used in the HYP procedure, 
at one-loop order of PT; we show that some results, previously known in the 
literature, are largely independent of how the projection is defined. In 
sect.~4 we provide a simple derivation of the Feynman rules of the temporal 
HYP link in time-momentum representation. A systematic study of the one-loop 
static self-energy  is described in sect.~5. In sect.~6 we collect our results 
for the improvement coefficients. Conclusions are drawn in sect.~7. For the 
sake of readability we strived to avoid as much as possible unessential 
technicalities. For this reason, algebraic details have been almost completely
confined to the appendices.

\section{Notation and basic definitions}
\subsection{HYP static action}

In order to set-up the notation, we review some basic formulae concerning
the static theory and the HYP link. Static quarks are represented by a pair 
of fermion fields $(\heavy,\aheavy)$, propagating forward and backward in 
time, respectively; their dynamics is governed by the lattice action 
\cite{Eichten:1989zv}
\begin{gather}
\label{EHaction}
  S^{\rm stat}[\heavy,\aheavy] = a^4\sum_x
  \left[\heavyb(x)\nabla_0^*\heavy(x) -
    \aheavyb(x)\nabla_0\aheavy(x)\right],
\end{gather}
where the forward and backward covariant derivatives
$\nabla_0,\,\nabla_0^*$ are defined via
\begin{align}
& \nabla_0\aheavy(x)=\frac{1}{a}\left[W_0(x)\aheavy(x+a\hat{0})
 -\aheavy(x)\right],  \nonumber \\
& \nabla_0^*\heavy(x)=\frac{1}{a}\left[\heavy(x)-W_0(x-a\hat{0})^{-1}
 \heavy(x-a\hat{0})\right].
\label{eq_latt_der}
\end{align}
The field $\heavy$($\heavyb$) can be thought of as the annihilator(creator) 
of a heavy quark. Similarly, $\aheavy$($\aheavyb$) creates(annihilates) a heavy
anti-quark. Each field is represented by a four-component Dirac vector, yet 
only half of the components play a dynamical r\^ole, owing to the static
projection constraints
\begin{align}
\label{constraints}
& P_\lp\heavy = \heavy\ ,\qquad \heavyb P_\lp = \heavyb\ ,\qquad
  P_\lp = \frac{1}{2}(\id+\gamma_0)\ ;  \nonumber \\
& P_\lm\aheavy = \aheavy\ ,\qquad \aheavyb P_\lm =
  \aheavyb\ ,\qquad P_\lm = \frac{1}{2}(\id-\gamma_0)\ .
\end{align}
\parbreak
The static action has a functional dependence upon the parallel transporter 
$W_0$, which is assumed to be represented in this paper by a temporal HYP 
link. According to the original definition \cite{Hasenfratz:2001hp}, the HYP
link is obtained through a three-step recursive smearing procedure, where
the original thin link is decorated with staples belonging to its 
surrounding hypercube, i.e.
\begin{alignat}{1}
\label{hyper1}
W_\mu(x) \equiv W^{(3)}_\mu(x) & = {\cP}_{\rm\scriptscriptstyle SU(3)}%
[(1-\alpha _{1})U_\mu(x) + \nonumber \\[1.3ex]%
& + \frac{\alpha _{1}}{6}\sum _{\pm \nu \neq \mu }W^{(2)}_{\nu ;\mu}(x)%
W^{(2)}_{\mu ;\nu}(x+a\hat{\nu })W^{(2)}_{\nu ;\mu}({x+a\hat{\mu }})%
^{\dagger }]\, , \\[2.0ex]
\label{hyper2}
W^{(2)}_{\mu ;\nu }(x) & = {\cP}_{\rm\scriptscriptstyle SU(3)}[(1-\alpha%
_{2})U_{\mu}(x)+ \nonumber \\[1.3ex]
& + \frac{\alpha _{2}}{4}\sum _{\pm \rho \neq \nu ,\mu }W^{(1)}_{\rho ;\nu%
\mu }(x)W^{(1)}_{\mu ;\rho\nu }(x+a\hat{\rho})W^{(1)}_{\rho ;\nu\mu }%
(x+a\hat{\mu})^{\dagger }]\, , \hfill \\[2.0ex]
\label{hyper3}
W^{(1)}_{\mu ;\nu\rho}(x) & = {\cP}_{\rm\scriptscriptstyle SU(3)}%
[(1-\alpha _{3})U_{\mu}(x)+ \nonumber \\[1.3ex]
& + \frac{\alpha_{3}}{2}\sum _{\pm \eta \neq \rho ,\nu ,\mu }%
U_{\eta}(x)\ U_{\mu}(x+a\hat{\eta })U_{\eta}(x+a\hat{\mu })^{\dagger }]\, ,
\end{alignat}
where $U_{-\mu}(x) = U^\dagger_\mu(x-a\hat\mu)$. In \Eq{hyper1} \( U_{\mu } \) 
denotes the fundamental gauge link and the index \( \nu  \) in \( W^{(2)}_{\mu ;\nu } \) 
indicates that the fat link at location \( x \) and direction \( \mu  \) is 
not decorated with staples extending in direction \( \nu . \) The decorated link 
\( W^{(2)}_{\mu ;\nu } \) is then constructed in \Eq{hyper2} with a modified APE 
blocking from another set of decorated links, where the indices \( \rho \, \nu  \) 
indicate that the fat link \( W^{(1)}_{\mu ;\rho \, \nu } \) in direction \( \mu  \) 
is not decorated with staples extending in the \( \rho  \) or \( \nu  \) 
directions. Finally, the decorated link \( W^{(1)}_{\mu ;\rho \, \nu } \) is 
constructed in \Eq{hyper3} from the original thin links with a modified APE 
blocking step where only the two staples orthogonal to \( \mu ,\, \nu  \) and 
\( \rho  \) are used. After each smearing step, the new fat link is projected onto 
SU(3). Popular choices of the smearing parameters are represented by
\begin{align}
\label{EHchoice}   \rm{EH}  \ :  & \hspace{2.0cm} (\alpha_1,\alpha_2,\alpha_3) \, = \, (\ \,0.0,\ \,0.0,\ \,0.0)\ ,  \\[1.5ex]
\label{HYPoaction} \rm{HYP1}\ :  & \hspace{2.0cm} (\alpha_1,\alpha_2,\alpha_3) \, = \, (   0.75,\ \,0.6,\ \,0.3)\ , \\[1.5ex]
\label{HYPtaction} \rm{HYP2}\ :  & \hspace{2.0cm} (\alpha_1,\alpha_2,\alpha_3) \, = \, (\ \,1.0,\ \,1.0,\ \,0.5)\ .
\end{align}

The APE action of ref.~\cite{DellaMorte:2005yc} is characterised by a 
parallel transporter, which is not projected onto SU(3). Therefore, the
action cannot be generated through a HYP link for any choice of the 
smearing parameters at a non-perturbative level. Nevertheless, the 
SU(3) projection is irrelevant at one-loop order of PT, as shown in 
next section. It follows that, as far as we are concerned here, the APE 
action is obtained from 
\begin{align}
\rm{APE}  \ :  & \hspace{2.0cm} (\alpha_1,\alpha_2,\alpha_3) \, = \, ( 1.0,\ \,0.0,\ \,0.0)\ .
\end{align}

\subsection{\Oa improvement of the static currents}

We adopt a SF topology where periodic boundary conditions (up to
a phase $\theta$ for the light quark fields) are set up on the
spatial directions and Dirichlet boundary conditions are imposed
on time at $x_0=0,T$. We also assume no background field in our
theory. The SF formalism has certain advantages, which
have been widely discussed in the literature; we refer the reader
to \cite{Sommer:2006sj} for an introduction to the subject. In
particular, it is here adopted since it allows for an easy 
on-shell formulation of the improvement problem.

\vspace{0.1cm}

In this paper, we are interested in the static axial and vector 
currents
\begin{align}
A_0^{\rm stat}(x) & = \overline{\psi}(x)\gamma_0\gamma_5 \heavy(x)\ , \\[1.5ex]
V_0^{\rm stat}(x) & = \overline{\psi}(x)\gamma_0 \heavy(x)\ .
\end{align}
The \Oa improvement of these operators with Wilson light 
and EH static quarks has been studied at one-loop order of PT
in \cite{Palombi:2007dt,Kurth:2000ki}. The general improvement
pattern is as follows:
\begin{align}
(A_{\rm\scriptscriptstyle R}^{\rm stat})_0 & = \ZAstat(1 + \bAstat am_{\rm q})(A_{\rm\scriptscriptstyle I}^{\rm stat})_0 \ , \\[1.5ex]
(V_{\rm\scriptscriptstyle R}^{\rm stat})_0 & = \ZVstat(1 + \bVstat am_{\rm q})(V_{\rm\scriptscriptstyle I}^{\rm stat})_0\ , 
\end{align}
with
\begin{alignat}{3}
(A_{\rm\scriptscriptstyle I}^{\rm stat})_0 & = A_0^{\rm stat} + a\cAstat \delta A_0^{\rm stat} \ , 
\qquad & \quad \delta A_0^{\rm stat}(x) & = \overline{\psi}(x)\gamma_j\gamma_5\frac{1}{2}\left(
\overleftarrow{\nabla}_j + \overleftarrow{\nabla}^*_j\right)\heavy(x)\ ; \\[1.5ex]
(V_{\rm\scriptscriptstyle I}^{\rm stat})_0 & = V_0^{\rm stat} + a\cVstat \delta V_0^{\rm stat} \ , 
\qquad & \quad \delta V_0^{\rm stat}(x) & = \overline{\psi}(x)\gamma_j\frac{1}{2}\left(\overleftarrow{\nabla}_j + 
\overleftarrow{\nabla}^*_j\right)\heavy(x)\ . 
\end{alignat}
The improvement coefficients $\cAstat$, $\cVstat$, $\bAstat$ and $\bVstat$
depend on the gauge coupling and are perturbatively expanded according to
\begin{alignat}{3}
c_{\rm\scriptscriptstyle X}^{\rm stat} & = c_{\rm\scriptscriptstyle X}^{{\rm stat}(1)}g_0^2 + {\rm O}(g_0^4)\ , 
\qquad & \qquad \  \\[1.5ex]
b_{\rm\scriptscriptstyle X}^{\rm stat} & = \frac{1}{2} + b_{\rm\scriptscriptstyle X}^{{\rm stat}(1)}g_0^2 + {\rm O}(g_0^4)\ ;
\qquad & \qquad {\rm X} = {\rm A,V}\ . 
\end{alignat}
We do not review explicitely the improvement conditions used in 
\cite{Palombi:2007dt,Kurth:2000ki} here, but perform the same 
perturbative analysis with HYP action and refer the reader to 
those papers for details. In particular, we extract the 
one-loop improvement coefficients $\cAstatone$ and $\bAstatone$ of 
the static axial current as  explained in 
appendix B of \cite{Kurth:2000ki}, and the one-loop improvement 
coefficients $\cVstatone$ and $\bVstatone$ of the static vector 
current according to Eqs.~(4.3)-(6.13) of 
\cite{Palombi:2007dt}. 
\section{${\rm SU}(3)$ projection at one-loop order of PT}
The ${\rm SU}(3)$ projector ${\cP}_{\rm\scriptscriptstyle SU(3)}$ 
is not uniquely defined and the resulting HYP link depends in 
principle upon the chosen definition. Here we follow 
\cite{Della Morte:2005yc}: if $A$ denotes a generic $3\times 3$ 
complex matrix, its projection $M$ onto SU(3) is defined by
\begin{equation}
\label{alphaproj}
M = \frac{X}{\sqrt[3]{(\det{X})}}\ , \qquad X = {\ A}({{A^\dagger A}})^{-1/2}\ ,
\end{equation}
i.e. the matrix $A$ is first unitarised, and then its determinant 
is rotated to one. Another very popular definition goes through 
the maximisation (minimisation) of the real (imaginary) part of 
${\rm tr}({M^\dagger A})$. The latter has been employed in particular 
in \cite{Lee:2002fj}, where two valuable propositions have been proved, 
which allow for a substantial simplification of perturbative 
calculations at one-loop order. The propositions concern the adjoint 
field of the HYP link, when this is expanded in power series of the 
adjoint (gluon) field of the fundamental gauge link. They read as 
follows:

\begin{itemize}
\item[1.]{\sl the linear term of the series is invariant under the action of 
${\cP}_{\rm\scriptscriptstyle SU(3)}$;}
\item[2.]{\sl the quadratic term is anti-symmetric in the gluon field.}
\end{itemize}

Although the proof proposed in \cite{Lee:2002fj} is based on the 
specific definition of the SU(3) projector as 
$\max_M\{{\rm Re\,tr}[M^\dagger A]\}$, the above statements hold as 
well if \Eq{alphaproj} is adopted instead. A proof of this is 
straightforward. To fix the notation, we define the gauge field 
$q_\mu^a(x)$ of the fundamental link via
\begin{equation}
U_\mu(x) = \exp\{ag_0q_\mu^a(x)T^a\}\ , 
\end{equation}
where $\{T^a\}_{a=1\dots 8}$ denotes an anti-hermitian representation 
of the Gell-Mann matrices. Following \cite{Lee:2002fj}, the unprojected 
APE-blocked links of Eqs.~(\ref{hyper1})--(\ref{hyper3}) are written 
in the general form
\begin{equation}
V_\mu(x) = \id\ + \ ag_0\sum_{\nu;y} f_{\mu\nu}(x,y)q_\nu(y) \ + \ 
 a^2g_0^2\sum_{\nu\rho;yz}h_{\mu\nu\rho}(x,y,z)q_\nu(y)q_\rho(z) \ + \ {\rm O}(g_0^3)\ ,
\end{equation}
with real vertices $f,h$ depending on the level of the HYP smearing and $q_\mu(x) = q_\mu^a(x)T^a$.
It should be observed that, since $V_\mu$ is neither a ${\rm SU}(3)$ nor a ${\rm U}(3)$ matrix, 
no gauge field can be associated with it. Projecting the link according to \Eq{alphaproj} leads to 
\begin{align}
\label{hypprojlink}
{V_\mu(x)}\left[{{V^\dagger_\mu(x)V_\mu(x)}}\right]^{-1/2} \ = \ \id & \ + \ ag_0\sum_{\nu;y} f_{\mu\nu}(x,y)q_\nu(y) \nonumber \\[1.1ex]
& \ + \ \frac{a^2g_0^2}{2}\sum_{\nu\rho;yz}h_{\mu\nu\rho}(x,y,z)[q_\nu(y),q_\rho(z)] \nonumber \\[1.1ex]
& \ + \ \frac{a^2g_0^2}{2}\sum_{\nu\rho;yz}f_{\mu\nu}(x,y)f_{\mu\rho}(x,z)q_\nu(y)q_\rho(z) \nonumber \\[1.5ex]
& \ + \ {\rm O}(g_0^3) \ , 
\end{align}
and
\begin{equation}
\label{hypdet}
\det\left\{{V_\mu(x)}\left[{{V^\dagger_\mu(x)V_\mu(x)}}\right]^{-1/2}\right\} \ = \ 1 \ + \ {\rm O}(g_0^3)\ .
\end{equation}
\Eq{hypdet} can been easily checked through an algebraic program, 
such as FORM \cite{Vermaseren:2000nd}. The projected link 
$\cP_{\rm\scriptscriptstyle SU(3)}[V_\mu(x)]$ is now in ${\rm SU}(3)$  
by construction. Consequently, we are allowed to introduce a 
smeared gauge field
\begin{equation}
\label{hypgauge}
\cP_{\rm\scriptscriptstyle SU(3)}[V_\mu(x)] = \exp\{ag_0v_\mu^a(x)T^a\}\ .
\end{equation}
As a remark we also note that, since the perturbative expansion 
of the staples generates gauge fields at all orders in PT, $v_\mu$ 
cannot be assumed as a linear function of $q_\mu$, and has to be 
perturbatively expanded in its turn, i.e.
\begin{equation}
v_\mu = v'_{\mu} \ + \ ag_0v''_{\mu} \ + \ {\rm O}(g_0^2) \ . 
\end{equation}
Accordingly, \Eq{hypgauge} reads
\begin{equation}
\label{hypexp}
\cP_{\rm\scriptscriptstyle SU(3)}[V_\mu(x)] = \id \ + \ ag_0v'_{\mu} \ + \ a^2g_0^2v''_{\mu} \ + \ \frac{a^2g_0^2}{2}v'_{\mu}v'_{\mu} \ + \ {\rm O}(g_0^3)\ . 
\end{equation}
Equating \Eq{hypprojlink} and \Eq{hypexp} order by order in the 
bare coupling allows to express the smeared gauge field in terms 
of the unsmeared one. In particular
\begin{alignat}{3}
{\rm O}(g_0^0): \qquad & & \id \ = \ & \id\ ,  \label{Lee0} \hskip 5.0cm \\[2.7ex]
{\rm O}(g_0^1): \qquad & & v'_{\mu} \ = \ & \sum_{\nu;y} f_{\mu\nu}(x,y)q_\nu(y)\ , \label{Lee1} \\[0.0ex]
{\rm O}(g_0^2): \qquad & & \ v''_{\mu} \ = \ & \frac{1}{2}\sum_{\nu\rho;yz}h_{\mu\nu\rho}(x,y,z)[q_\nu(y),q_\rho(z)] \ . \label{Lee2}
\end{alignat}
Eqs.~(\ref{Lee1})--(\ref{Lee2}) correspond precisely to propositions 
1 and 2 of \cite{Lee:2002fj}. It is worth noting that tadpole 
diagrams associated with \Eq{Lee2} do not contribute to any 
one-loop perturbative calculation in absence of a background field, 
although they are ${\rm O}(g_0^2)$. Indeed, the commutator
$[T^a,T^b] = -f^{abc}T^c$ is anti-symmetric in $a\leftrightarrow b$, 
while the Wick contraction $\langle q^a q^b\rangle$ is symmetric. 
Nevertheless, one should avoid the conclusion that tadpoles do not 
contribute at all, because terms such as $v'_\mu(x)v'_\mu(x)$ have 
to be always considered. Moreover, the vanishing of the above-mentioned
contributions is not anymore true in presence of a background 
field, where PT becomes more cumbersome.

To summarise, Eqs.~(\ref{Lee1})--(\ref{Lee2}) show that the smeared 
gauge field can be taken as a linear function of the unsmeared one
at one-loop in PT and zero background field. The effect of the 
${\rm SU}(3)$ projection can be disregarded in the linear term, and 
non-linear contributions should be discarded when deriving the 
Feynman rules of the HYP link.

\section{Feynman rules in time-momentum representation}
Translational invariance along time is broken within the
SF formalism. This fact complicates perturbation theory, 
which is usually worked out in momentum space, and makes
it natural to adopt a mixed approach, known as 
{\it time-momentum representation}, where only the 
spatial coordinates are Fourier transformed. Accordingly,
tree-level propagators and interaction vertices are 
functions of time and spatial momenta.


We derive the Feynman rules in time-momentum representation
for the temporal component of the HYP link according to two 
different and independent procedures. The first derivation 
makes use of the results obtained in full momentum 
space \cite{Hasenfratz:2001tw} and 
arrives at time-momentum representation by inverse Fourier 
transform in time. This is possible in presence of homogeneous
Dirichlet boundary conditions, which allow to continue the
SF periodically to all times with no non-trivial terms at the
boundaries. The second derivation follows by direct 
construction upon representing the HYP smearing as a lattice 
differential operator acting linearly on the fundamental gauge 
fields. In this section we report only the first derivation; 
the second one, which is quite lengthy, is sketched in appendix
A. For the sake of completeness, we report in appendix B
the Feynman rules in time-momentum representation for the 
spatial components of the HYP link, which may be useful for 
some applications of HQET (e.g. see ref.~\cite{Guazzini:2007bu})
and for dynamical n-HYP \cite{Hasenfratz:2007rf}.


In order  to fix the notation, we define the smeared gauge fields of 
the three levels of the HYP procedure according to
\begin{align}
W^{(k)}_\mu(x) & = \exp\{ag_0B^{(k)}_\mu(x)\}\ , \qquad \qquad k = 1,2,3\ .
\end{align}
Under the assumption of periodic boundary conditions in space
and homogenous Dirichlet ones in time without nontrivial boundary
terms in the action, the gauge field of the 
fundamental (and smeared) link admits a 4-dimensional Fourier 
transform
\begin{equation}
q_\mu(x) = \frac{1}{L^4}\sum_p \re^{ipx}\re^{i\frac{a}{2}p_\mu}
\tilde q_\mu(p)\ , \qquad
-\frac{\pi}{a} \le p_\mu = \frac{2\pi}{L}n_\mu < \frac{\pi}{a}\ ,
\end{equation}
where $L$ is the toroidal extension of the space-time dimensions and
the additional phase shift in direction $\hat\mu$ is due to the very
definition of the gauge fields, which are supposed to live between
neighboring lattice points. Factorising the loop-sums over the spatial
directions allows to express the gauge field in time-momentum 
representation in terms of the fully Fourier-transformed one, i.e.
\begin{align}
q_0(x) & = \frac{1}{L^3}\sum_\bp \re^{i\bp\bx}\left[\frac{1}{L}\sum_{p_0}
\re^{ip_0x_0}\re^{i\frac{a}{2}p_0}\tilde q_0(p)\right] = 
\frac{1}{L^3}\sum_\bp \re^{i\bp\bx} \tilde q_0(x_0;\bp)\ , \\[1.5ex]
q_k(x) & = \frac{1}{L^3}\sum_\bp \re^{i\bp\bx}\re^{i\frac{a}{2}p_k}\left[\frac{1}{L}\sum_{p_0}
\re^{ip_0x_0}\tilde q_k(p)\right] = 
\frac{1}{L^3}\sum_\bp \re^{i\bp\bx} \re^{i\frac{a}{2}p_k} \tilde q_k(x_0;\bp)\ .
\end{align}
By direct comparison, it follows
\begin{align}\label{e:FAT}
\tilde q_0(x_0;\bp) & = \frac{1}{L}\sum_{p_0}
\re^{ip_0x_0}\re^{i\frac{a}{2}p_0}\tilde q_0(p)\ , \\[1.2ex]
\tilde q_k(x_0;\bp) & = \frac{1}{L}\sum_{p_0}
\re^{ip_0x_0}\tilde q_k(p)\ . 
\end{align}
We now consider the Feynman rules in momentum representation \cite{Hasenfratz:2001tw}, 
\begin{equation}
\label{eq:decomp}
\tilde B_\mu^{(3)}(p) = \sum_\nu f_{\mu\nu}(p)\tilde q_\nu(p) \ + \ {\rm O}(g_0)\ , 
\end{equation}
where
\begin{align}
f_{\mu\nu}(p) & = \delta_{\mu\nu}\left[1 - \frac{\alpha_1}{6}\sum_\rho (a^2\hat p_\rho^2) \Omega_{\mu\rho}(p)
\right] + \frac{\alpha_1}{6}(a\hat p_\mu)(a\hat p_\nu)\Omega_{\mu\nu}(p) \ , \\[1.5ex]
\Omega_{\mu\nu}(p) & = 1 + \alpha_2(1+\alpha_3) - \frac{\alpha_2}{4}(1+2\alpha_3)a^2(\hat p^2 - \hat p_\mu^2 - 
\hat p_\nu^2) + \frac{\alpha_2\alpha_3}{4}\prod_{\eta\ne\mu,\nu}a^2\hat p_\eta^2 \ . 
\end{align}
Since we are interested in the static propagator, 
we focus on $B_0^{(3)}$, for which we get
\begin{equation}
\tilde B_0^{(3)}(x_0;\bp) = \frac{1}{L}\sum_{p_0}
\re^{ip_0x_0}\re^{i\frac{a}{2}p_0}\left[ f_{00}(p)\tilde q_0(p) +  \sum_{k=1}^3 f_{0k}(p)\tilde q_k(p)
\right]\ , 
\end{equation}
We then observe that
\begin{align}
f_{00}(p) & = 1 - \frac{\alpha_1}{6}\sum_{k=1}^3 (a^2\hat p_k^2) \Omega_{0k}(p)\ , \\[1.5ex]
\Omega_{0k}(p) & =  1 + \alpha_2(1+\alpha_3) - \frac{\alpha_2}{4}(1+2\alpha_3)(a^2\hat p_j^2 + a^2\hat p_l^2) \ + \nonumber \\[1.2ex]
& + \frac{\alpha_2\alpha_3}{4}(a^2\hat p_j^2)(a^2\hat p_l^2) \ , \qquad (\hat j,\hat l) \perp \hat k\ , \\[1.5ex]
f_{0k}(p) & = \frac{\alpha_1}{6}(a\hat p_0)(a\hat p_k) \Omega_{0k}(p) \ .
\end{align}
The above equations show in particular that $\Omega_{0k}$ 
and $f_{00}$ depend only upon the spatial components of the 
Fourier momentum. Therefore, 
\begin{align}
\tilde B_0^{(3)}(x_0;\bp) & = f_{00}(\bp)\tilde q_0(x_0;\bp) \ + \nonumber \\[1.2ex] 
& + \frac{\alpha_1}{6}\sum_{k=1}^3 (a\hat p_k)
\Omega_{0k}(\bp)\frac{1}{L}\sum_{p_0}
\re^{ip_0x_0}\re^{i\frac{a}{2}p_0}\ (a\hat p_0)\tilde q_k(p) \ + \ {\rm O}(g_0)\ .
\end{align}
The second term on the right hand side of the previous 
equation can be easily worked out, %
and we finally obtain
\begin{equation}
\tilde B_0^{(3)}(x_0;\bp) = f_{00}(\bp)\tilde q_0(x_0;\bp) - \frac{i\alpha_1}{6}\sum_{k=1}^3 (a\hat p_k)
\Omega_{0k}(\bp)\partial_0\tilde q_k(x_0;\bp) \ .
\end{equation}
A very compact way of writing the above expression is 
by introducing an effective vertex and auxiliary 
indices, i.e.
\begin{equation}
\tilde B_0^{(3)}(x_0;\bp) = \sum_{i=0}^6 V_{0;i}(\bp)\tilde q_{\mu(i)}(x_0+as(i);\bp)\ , 
\end{equation}
all collected in Table~\ref{tab:Feynman}. Comparing the latter 
with Table~4 of \cite{Della Morte:2005yc} shows that the Feynman 
rules of the HYP link resemble closely those of the APE one, from
which they differ by the presence of a form factor $\Omega_{0k}$. 
\begin{table}[!t]
  \begin{center}
    \vbox{\vskip 0.2cm}
    \begin{tabular}{cccc}
      \hline\hline\\[-2.0ex]
      $i$ \ \ \ & $\mu(i)$ & $s(i)$ & \ \ \ $V_{0;i}(\bp)$  \\ \\[-2.0ex]
      \hline\\[-2.0ex]
      0 \ \ \ & 0 & 0 & \ \ \ $1 - \frac{\alpha_1}{6}\sum_{k=1}^3a^2\hat p_k^2 \Omega_{0k}(\bp)$ \\[1.5ex]
      1,2,3 \ \ \ & $i$ & 0 & $ \ \ \ + \frac{i\alpha_1}{6}a\hat p_{\mu(i)}\Omega_{0{\mu(i)}}(\bp)$ \\[1.5ex]
      4,5,6 \ \ \ & $i-3$ & 1 & \ \ \ $ - \frac{i\alpha_1}{6}a\hat p_{\mu(i)}\Omega_{0\mu(i)}(\bp)$ \\[1.5ex] 
      \hline\hline
    \end{tabular}
    \caption{{Feynman rules of the temporal HYP link in time-momentum representation.}\label{tab:Feynman}}
  \end{center}
  \vskip -0.5cm
\end{table}

\section{The static self-energy at one-loop order in perturbation theory}
The binding energy $E_{\rm stat}$ of a static-light meson controls the 
exponential decay rate of the associated two-point function. It depends 
upon the choice of the parallel transporter and diverges linearly in
the continuum limit. As such, it can be perturbatively expanded according 
to
\begin{equation}
\label{Estat}
E_{\rm stat} \ \sim \ E_{\rm self} + {\rm O}(a^0) \sim \frac{1}{a}e^{(1)}
g_0^2 + \dots \ . 
\end{equation}
The coefficient $e^{(1)}$ represents an ultraviolet property of 
the static action. Accordingly, it is insensitive to the specific 
correlation function from which it is measured, as well as to the 
choice of the boundary conditions. 

\vspace {0.1cm}

\begin{figure}[!t]
  \vskip -1.2cm
  \begin{center}
    \vskip 1.5cm
    \hskip -0.2cm \epsfig{figure=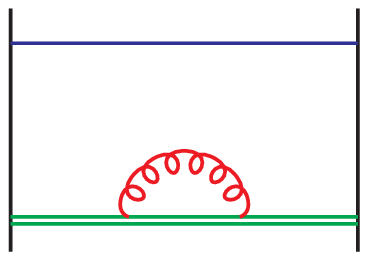, width=2.2 true cm}\hskip 2.0cm
    \epsfig{figure=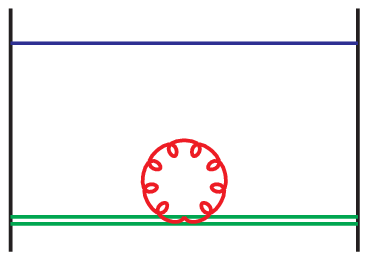, width=2.2 true cm}
  \end{center}
  \vskip -0.4cm
  \caption{\small Feynman diagrams contributing to the static self-energy at 
    one-loop order of PT. Euclidean time goes from left to right. Single (double)
    lines represent relativistic (static) valence quarks.     
  }
  \label{fig:Feynman}
\end{figure}

In appendix A of ref.~\cite{DellaMorte:2005yc}, $e^{(1)}$ is defined from 
the SF boundary-to-boundary correlator $f_1^{\rm hh}$, made of two
static quarks propagating across the bulk region. Here we adopt a different
definition and extract $e^{(1)}$ from the boundary-to-boundary correlator
$f_1^{\rm stat}$, made of one static and one relativistic propagator\footnote{For
the original definition of $f_1^{\rm stat}$, see Eq.~(3.23) of \cite{Kurth:2000ki}.}, 
i.e.
\begin{equation}
\label{eonedef}
  e^{(1)} = - \lim_{a/L\to 0} \frac{a}{L}\frac{f_{1,\,\rm\scriptscriptstyle self}^{{\rm\scriptscriptstyle stat}(1)}}{f_1^{{\rm\scriptscriptstyle stat}(0)}}\ .
\end{equation}
where $f_{1,\,\rm\scriptscriptstyle self}^{{\rm\scriptscriptstyle stat}(1)}$ denotes 
the  sum of the two Feynman diagrams depicted in Fig.~\ref{fig:Feynman}, 
with the light-quark line kept at tree-level. Evaluating \Eq{eonedef} leads to the 
expression
\begin{equation}
\label{eoneres}
e^{(1)} = \sum_{k_1k_2k_3=0}^2 e^{(1)}_{k_1k_2k_3}\alpha_1^{k_1}\alpha_2^{k_2}\alpha_3^{k_3}\ , 
\end{equation}
with non-zero coefficients $e^{(1)}_{k_1k_2k_3}$ collected in Table~\ref{fig:statcoefs}.
A derivation of \Eq{eoneres} is reported in appendix \ref{app:C}. The static self-energy is a 
multivariate polynomial of $(\alpha_1,\alpha_2,\alpha_3)$. 
The structure of the polynomial is triangular, i.e. the only non-vanishing contributions 
are at $0 \le k_3 \le k_2 \le k_1 \le 2$. This is an obvious consequence of how the three
steps of the HYP smearing are defined in Eqs.~(\ref{hyper1})--(\ref{hyper3}). 

It is also worth noting that higher order perturbative corrections to \Eq{Estat} have 
the same functional form (but different coefficients) as \Eq{eoneres}, since no HYP 
links appear in the light quark action, and therefore dynamical quark loops are 
directly connected to thin gluons only. This is not true anymore if the HYP link 
(or variants of it) enters the covariant derivatives of the Wilson action, such as 
in ref.~\cite{Hasenfratz:2007rf}. In this case, dynamical quark loops may have a 
functional impact on the static self-energy. 

A numerical minimisation of the coefficient $e^{(1)}$ as a function of the smearing
parameters is easily performed. To this aim we used a MATLAB routine. The global minimum 
is found for a choice $(\alpha_1^*,\alpha_2^*,\alpha_3^*)$ given by
\begin{equation}
\label{Emin}
e^{(1)}_{\rm min} = e^{(1)}|_{(\alpha_1^*,\alpha_2^*,\alpha_3^*)} = 0.03520(1)\ , \qquad 
(\alpha_1^*,\alpha_2^*,\alpha_3^*) = (1.0,\,0.9011,\,0.5196)\ .
\end{equation}

\begin{figure}[!t]
  \begin{center}
    \epsfig{file=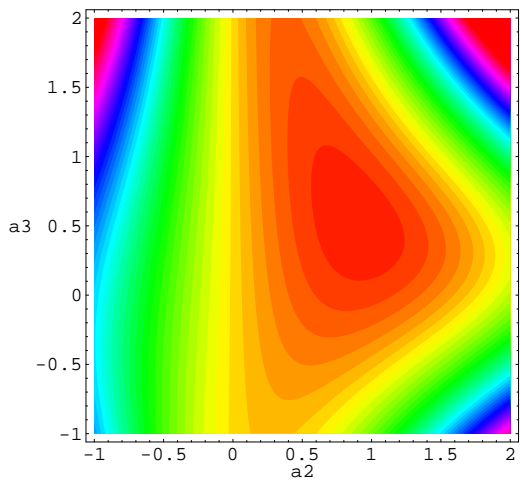,angle=0.0,width=0.45\textwidth}
    \caption{Countour plot of the coefficient $e^{(1)}$ as a function of 
      $\alpha_2$ and $\alpha_3$ at $\alpha_1=1.0$. The colour spectrum 
      goes from violet (higher values) to red (lower values).\label{fig:contour}}
  \end{center}
\end{figure}

A two-dimensional contour plot of $e^{(1)}$ as a function of $\alpha_2$ and $\alpha_3$ at
$\alpha_1=1.0$ is shown in Fig.~\ref{fig:contour}. Remarkably, the perturbative minimum 
is obtained with a choice of the smearing parameters which is very close to the HYP2 
definition, see \Eq{HYPtaction}, found via a numerical minimisation performed at a 
non-perturbative level \cite{DellaMorte:2005yc,Sommer:private}. The closeness of the 
perturbative and non-perturbative minima suggests that the static self-energy is dominated 
by PT, as intuitively expected. A perturbative comparison between \Eq{Emin} and HYP2 
is also possible: we find $e^{(1)}_{\rm\scriptscriptstyle HYP2} = 0.03544(1)$, which  
is slightly higher than the result reported in Table~1 of \cite{DellaMorte:2005yc} and 
differs by $0.7\%$ from the perturbative minumum. 

\section{Determination of the improvement coefficients}

Once the Feynman rules of the HYP link are known in
time-momentum representation, the extraction of the
improvement coefficients of the static axial and 
vector currents at one-loop order of PT can be 
performed along the lines of refs.~\cite{Palombi:2007dt,Kurth:2000ki}. 
As already warned in sect.~2, we do not review
here the improvement conditions (and related perturbative 
equations)  used in those papers,
since our analysis has nothing new to add from 
a methodological point of view%
\footnote{However, we observe that, while reproducing the calculation of $\bAstatone$
with EH action, we have found a little mistake in Eq.~(B.20) of
\cite{Kurth:2000ki}. This equation should be replaced by
\begin{equation}
\tilde X^{(1)} = (1 + \bAstatzero am_{\rm q}^{(0)})\left[ X^{(1)} + 
Z_{\rm\scriptscriptstyle A,lat}^{\rm stat (1)}X^{(0)}\right] + \bAstatzero
am_{\rm q}^{(1)}X^{(0)}\ . 
\end{equation}
and the correct value of the improvement coefficient is $\bAstatone = 0.0041(4)$ 
with EH action and $\bAstatone = 0.0925(6)$ with APE action. Note that the latter 
differs from Eq.~(3.8) of \cite{DellaMorte:2005yc}, which has been determined 
using the wrong value of $\bAstatone[{\rm EH}]$ as an input. Note also that 
all these values for $\bAstatone$ are numerically rather small. The mistake 
has been discussed with and recognised by the authors of refs.~\cite{Kurth:2000ki,DellaMorte:2005yc}.}. 
Instead, the reader is referred to appendix B of \cite{Kurth:2000ki}
as regards the improvement coefficients $\cAstatone$ 
and $\bAstatone$, and to Eqs.~(4.3)-(6.13) of
\cite{Palombi:2007dt} as for $\cVstatone$ and 
$\bVstatone$.

\vspace{0.1cm}

Upon this premises, we collect our 
results in  Table~\ref{tab:imprcoefs}. It is understood
 that all the conditions employed define 
the improvement coefficients up to \Oa terms and
need to be extrapolated to the continuum limit. 

\vspace{0.1cm}

\begin{center}
  \begin{figure}[!t]
    \begin{center}
      \begin{tabular}{p{0.45\textwidth}p{0.45\textwidth}}
        \epsfig{file=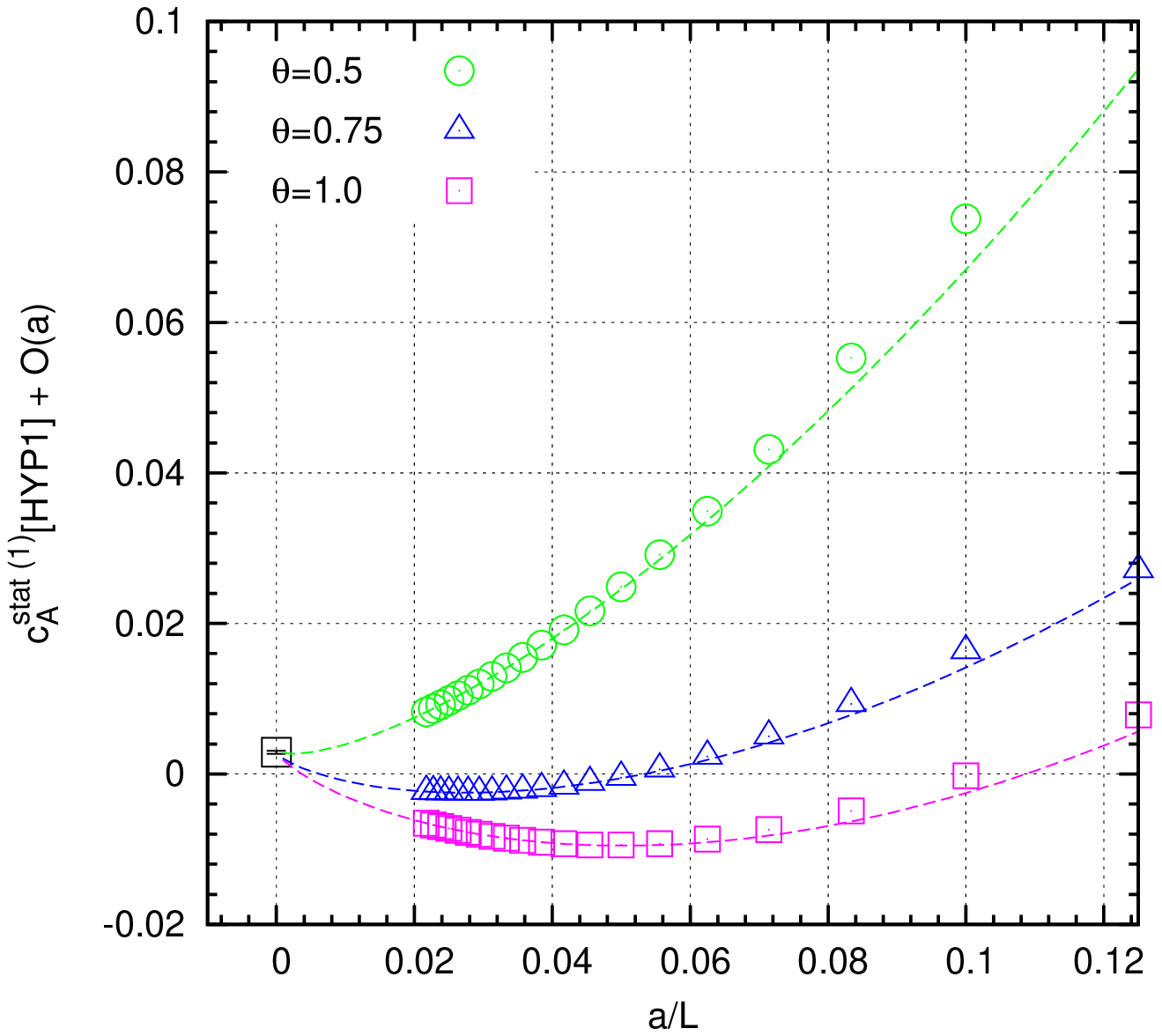,angle=0.0,width=0.45\textwidth}
        &
        \epsfig{file=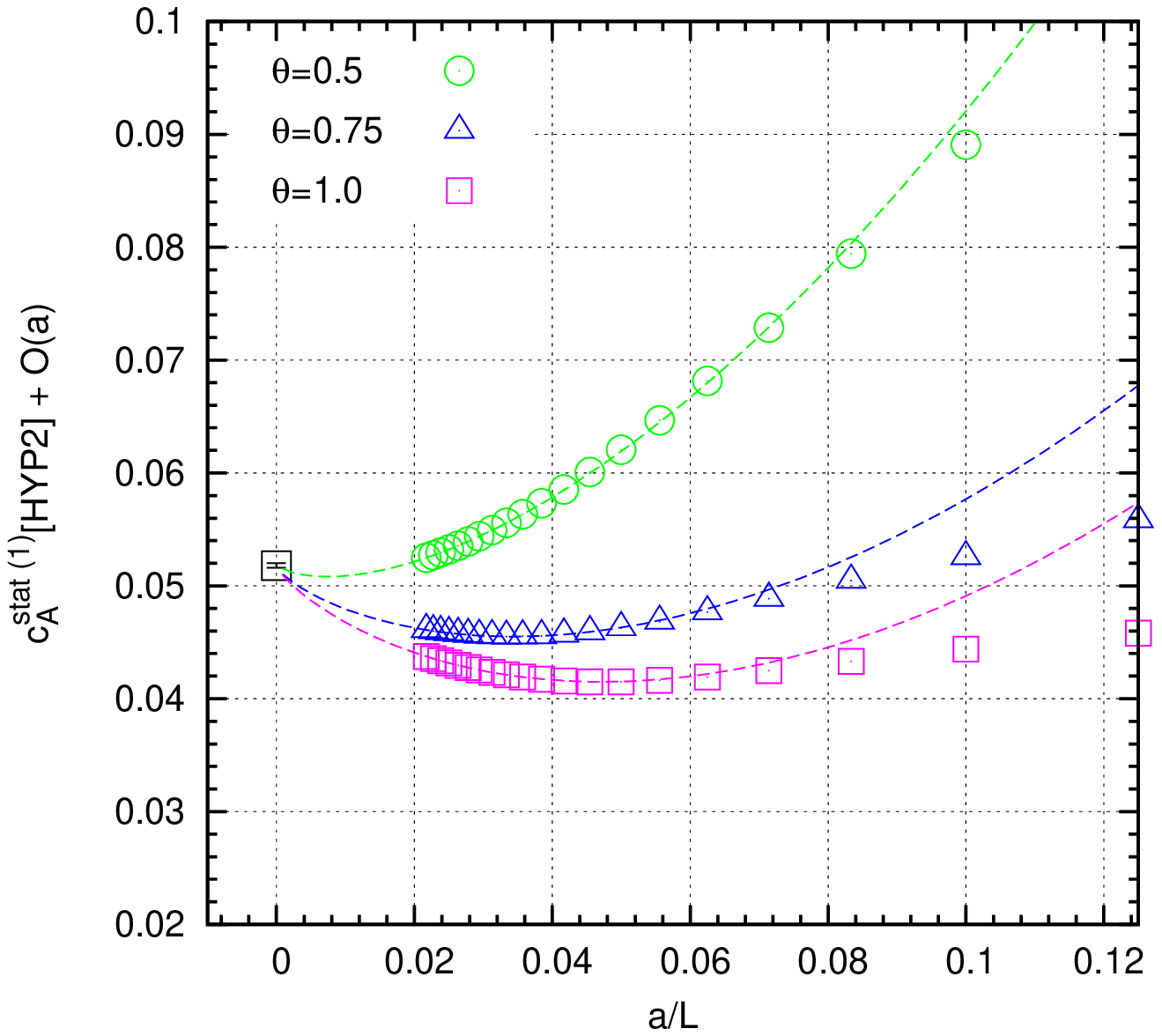,angle=0.0,width=0.45\textwidth}
      \end{tabular}
      \caption{Continuum extrapolation of the improvement coefficient $\cAstatone$ 
        with HYP1 (left) and HYP2 (right) static discretisation. On each plot, the 
        three curves refer to independent determinations at $\theta = 0.5$ (circles), 
        $\theta = 0.75$ (triangles) and $\theta=1.0$ (squares).\label{fig:cAstatHYP}}
      
      \vskip 0.5cm
      
      \begin{tabular}{p{0.45\textwidth}p{0.45\textwidth}}
        \epsfig{file=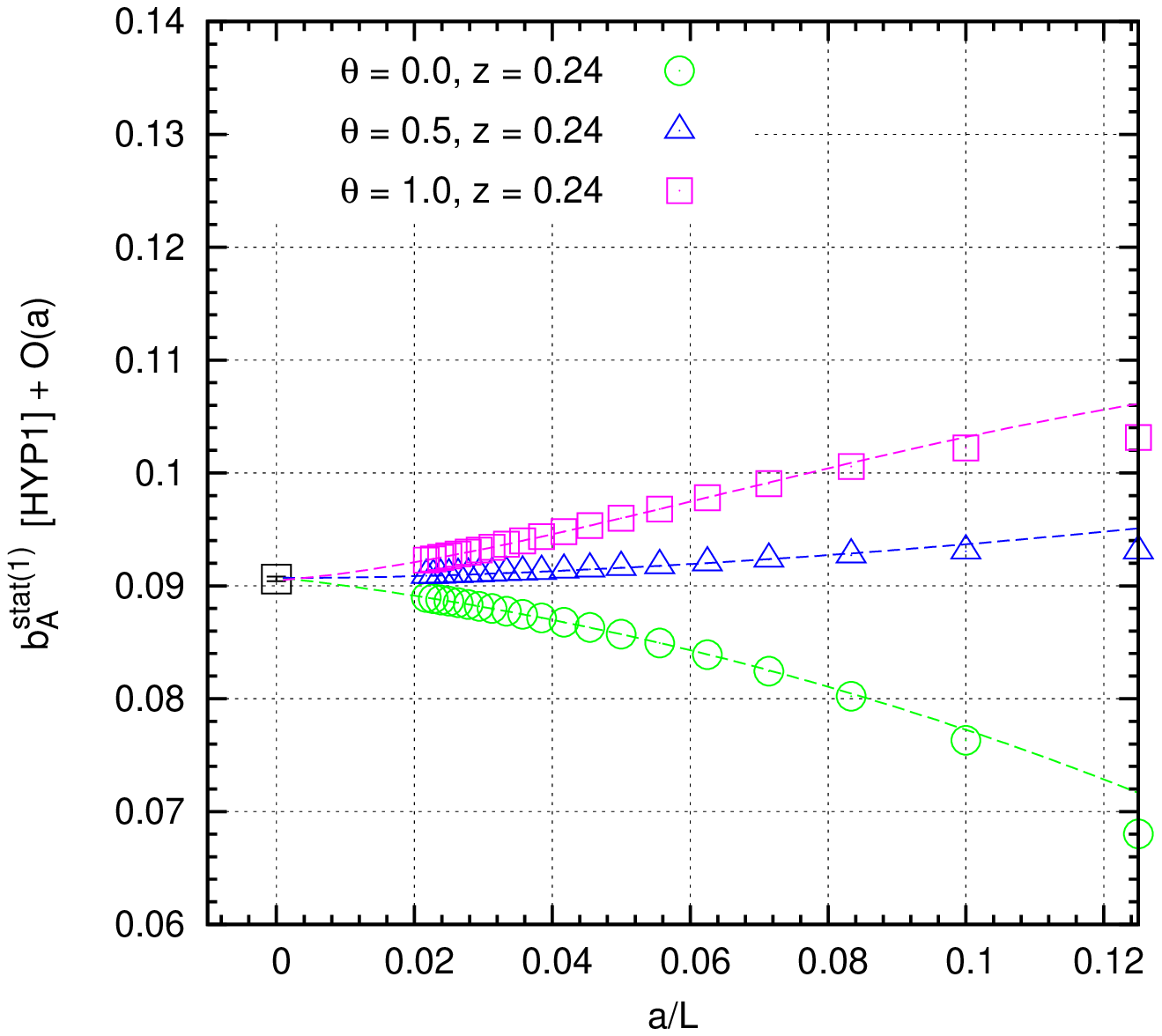,angle=0.0,width=0.45\textwidth}
        &
        \epsfig{file=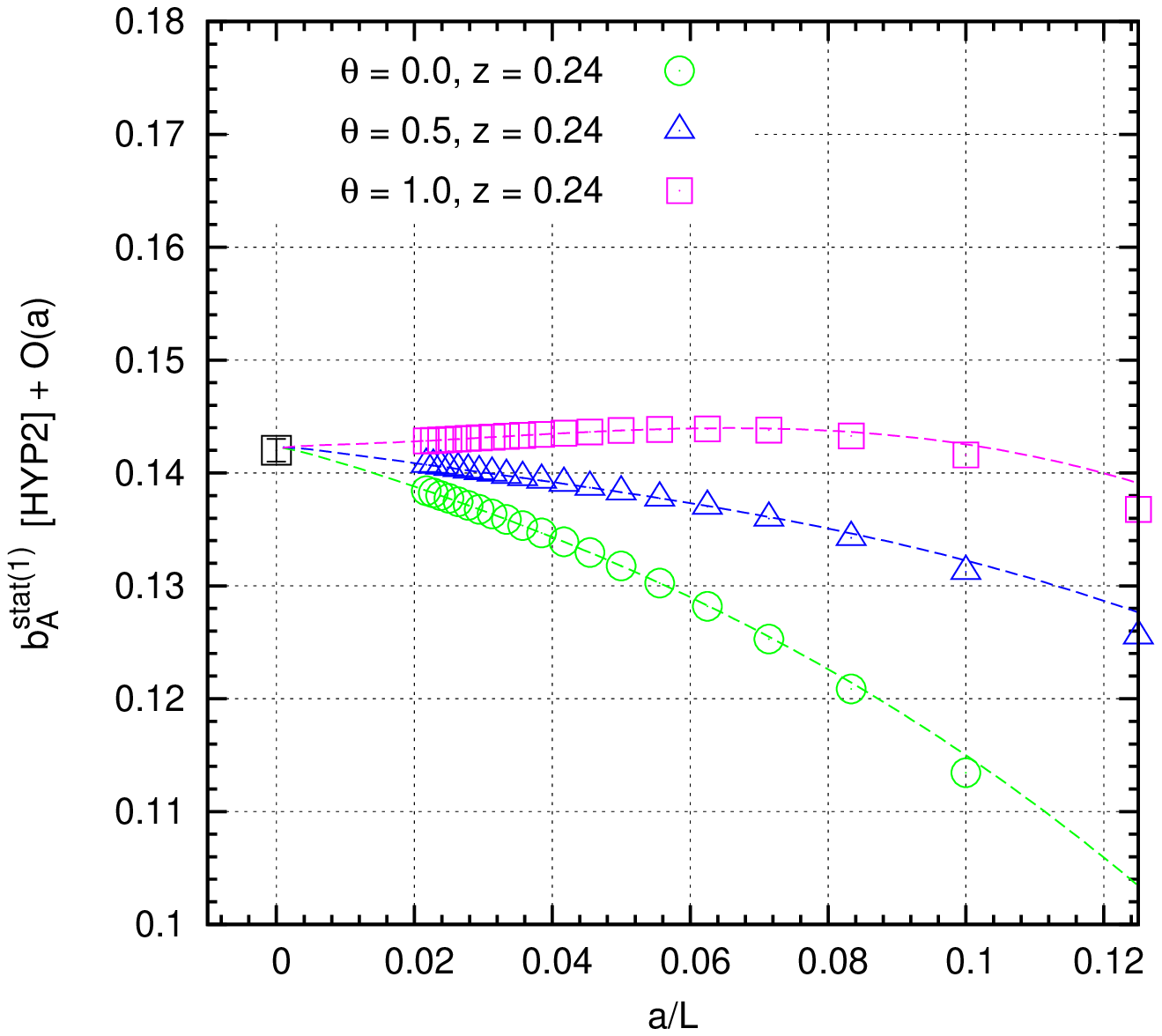,angle=0.0,width=0.45\textwidth}
      \end{tabular}
      \caption{Continuum extrapolation of the improvement coefficient $\bAstatone$ 
        with HYP1 (left) and HYP2 (right) static discretisation. On each plot, the 
        three curves refer to independent determinations at $\theta=0.0$ (circles), 
        $\theta = 0.5$ (triangles) and $\theta=1.0$ (squares).\label{fig:bAstatHYP}}
    \end{center}
  \end{figure}
\end{center}

For what concerns the axial current, we obtain the coefficient 
$\cAstatone$ from three independent improvement conditions, i.e. 
at $\theta\in\{0.5,0.75,1.0\}$ and renormalised quark mass 
$z=Lm_{\rm\scriptscriptstyle R}=0.0$. Having determined $\cAstatone$, 
we obtain $\bAstatone$ from three new independent conditions, i.e at 
$\theta\in\{0.0,0.5,1.0\}$\footnote{It is not possible to obtain 
$\cAstatone$ at $\theta=0$, since the operator $\delta A_0^{\rm stat}$ 
vanishes at tree-level for this particular choice of the $\theta$-angle.} 
and $z=0.24$. Continuum approach is shown in 
Figs.~\ref{fig:cAstatHYP}--\ref{fig:bAstatHYP}.

\vspace{0.1cm}

\begin{center}
  \begin{figure}[!t]
    \begin{center}
      \begin{tabular}{p{0.45\textwidth}p{0.45\textwidth}}
        \epsfig{file=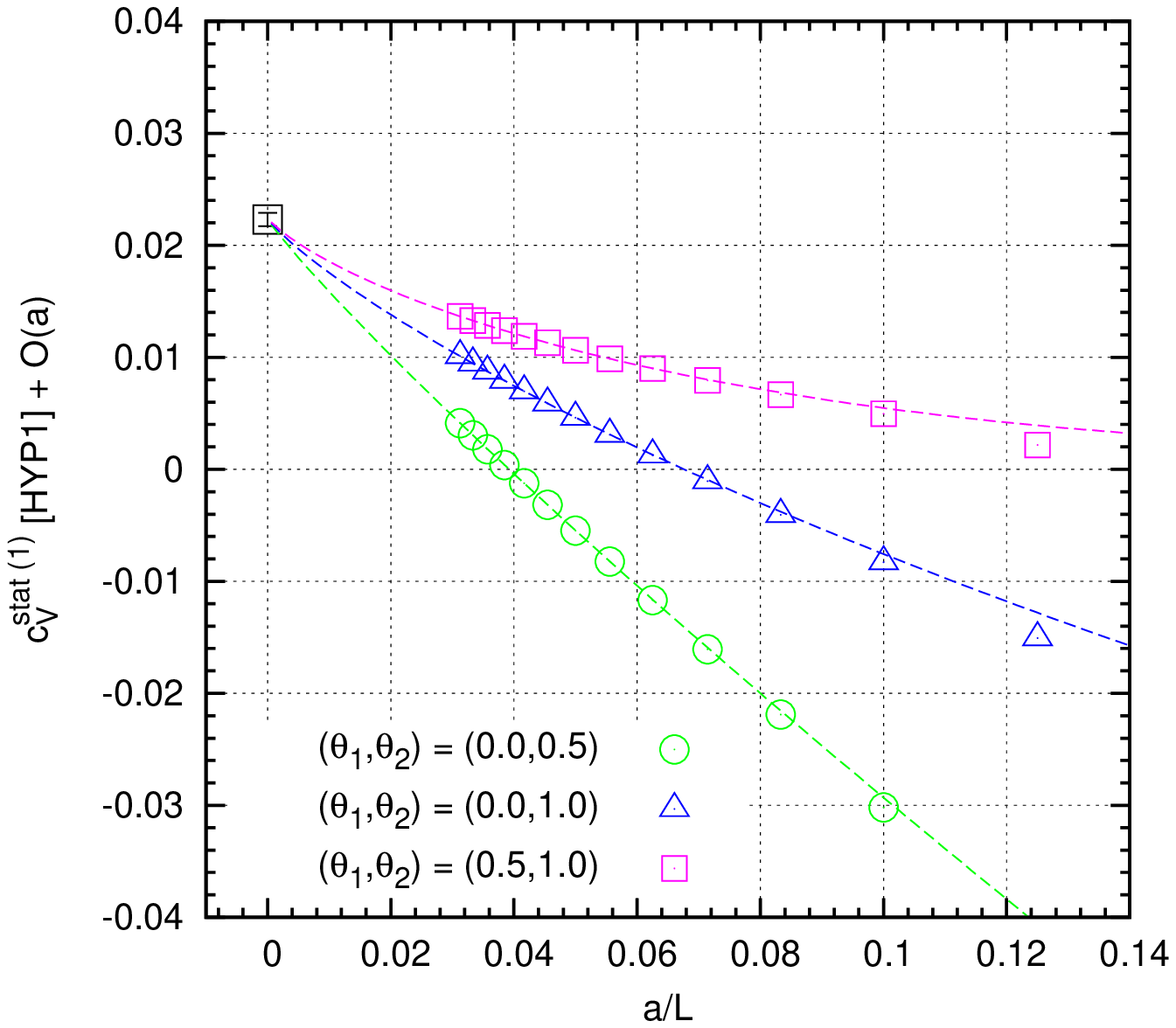,angle=0.0,width=0.45\textwidth}
        &
        \epsfig{file=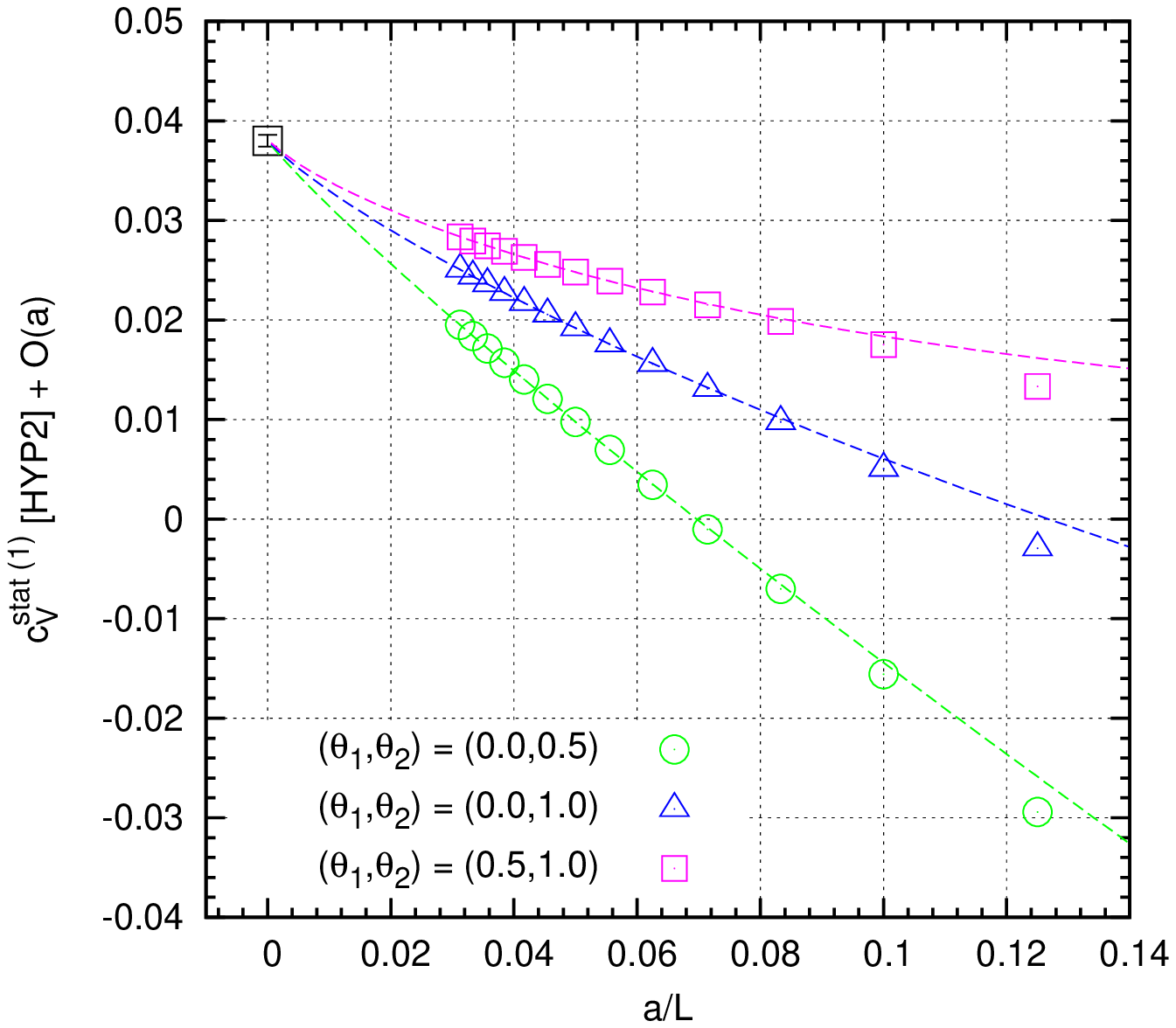,angle=0.0,width=0.45\textwidth}
      \end{tabular}
      \caption{Continuum extrapolation of the improvement coefficient $\cVstatone$ 
        with HYP1 (left) and HYP2 (right) static discretisation. On each plot, the 
        three curves refer to independent determinations at $(\theta_1,\theta_2) = (0.0,0.5)$ 
        (circles), $(\theta_1,\theta_2)=(0.0,1.0)$ (triangles) and 
        $(\theta_1,\theta_2)=(0.5,1.0)$ (squares).\label{fig:cVstatHYP}}
      
      \vskip 0.5cm
      
      \begin{tabular}{p{0.45\textwidth}p{0.45\textwidth}}
        \epsfig{file=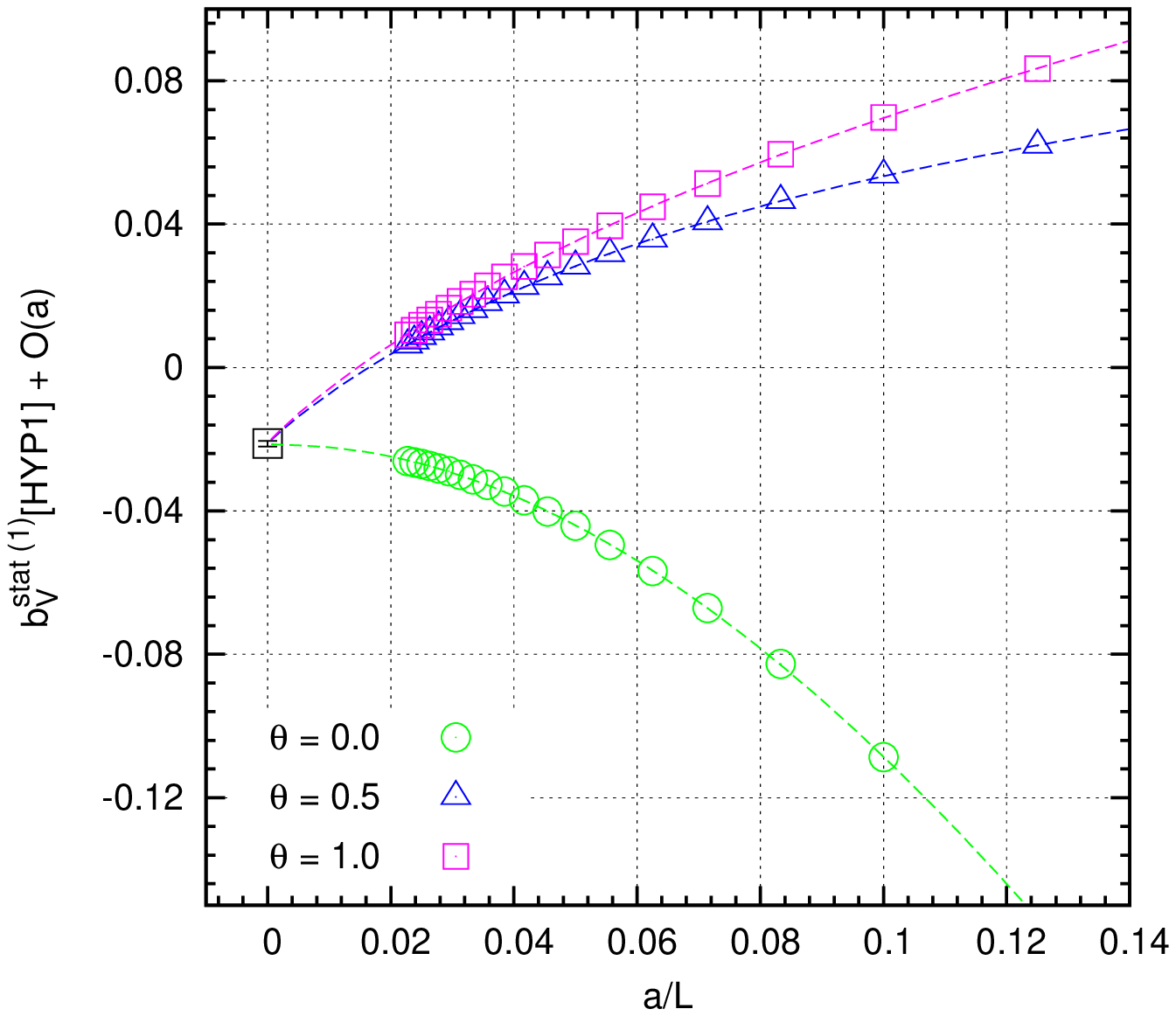,angle=0.0,width=0.45\textwidth}
        &
        \epsfig{file=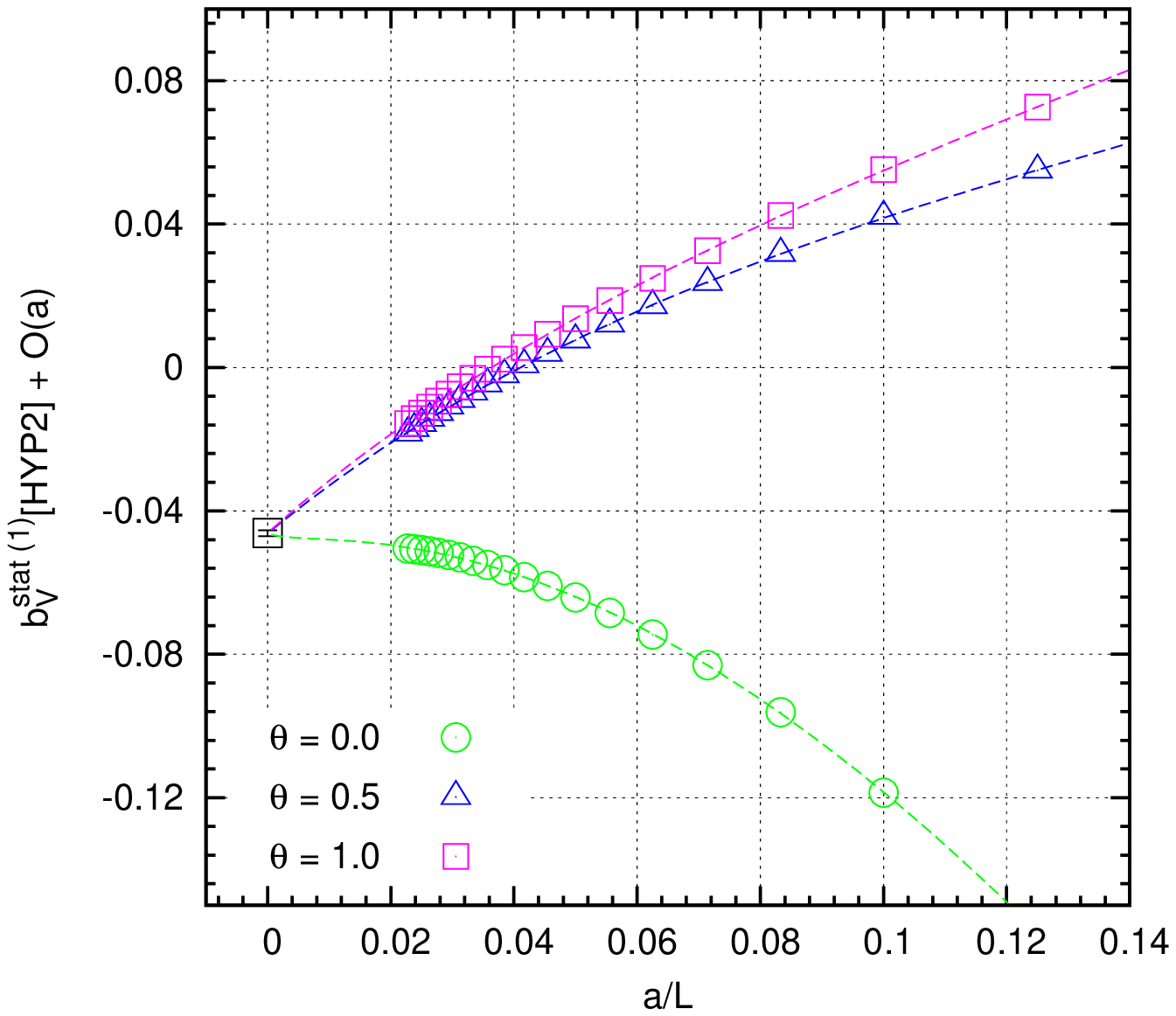,angle=0.0,width=0.45\textwidth}
      \end{tabular}
      \caption{Continuum extrapolation of the improvement coefficient $\bVstatone$ 
        with HYP1 (left) and HYP2 (right) static discretisation. On each plot, the 
        three curves refer to independent determinations at $\theta=0.0$ (circles), 
        $\theta = 0.5$ (triangles) and $\theta=1.0$ (squares).\label{fig:bVstatHYP}}
    \end{center}
  \end{figure}
\end{center}

As for the vector current, we obtain the coefficient $\cVstatone$ from three 
different implementations of the axial Ward Identity, corresponding to 
pairs of $\theta$-angles $(\theta_1,\theta_2) \in \{(0.0,0.5),(0.0,1.0),(0.5,1.0)\}$, 
and the coefficient $\bVstatone$ from the ratio of a three-point correlator of 
the static vector current at renormalised quark masses $(z_1,z_2) = (0.0,0.24)$ and 
three values of the $\theta$-angle, viz. $\theta\in\{0.0,0.5,1.0\}$. 
Convergence to the continuum limit is shown in 
Figs.~\ref{fig:cVstatHYP}--\ref{fig:bVstatHYP}.

\vspace{0.1cm}

In all cases, different definitions lead to a consistent 
continuum limit. The uncertainty on the final numbers has been 
estimated as the maximal difference of the continuum extrapolations 
corresponding to independent definitions. Data fits have been 
performed according to the asymptotic expansion
\begin{equation}
C\left(\frac{a}{L}\right) = C + A_1\left(\frac{a}{L}\right)
+ B_1\left(\frac{a}{L}\right)\log\left(\frac{a}{L}\right) + A_2\left(\frac{a}{L}\right)^2
+ B_2\left(\frac{a}{L}\right)^2\log\left(\frac{a}{L}\right) + \dots
\end{equation}
with $C = \cAstatone, \bAstatone, \cVstatone, \bVstatone$. 
\begin{center}
  \begin{table}[!t]
    \begin{tabular}{p{0.45\textwidth}p{0.45\textwidth}}
      \small
      \vskip 0.0cm
      \begin{center}
        \vskip -0.6cm
        \begin{tabular}{cc}
          \hline\hline\\[-2.0ex]
          $[i,j,k]$ & $e^{(1)}_{ijk}$  \\ \\[-2.0ex]
          \hline\\[-2.0ex]
          $[0,0,0]$ & $\phantom{-}0.168487(1)$  \\[0.2ex]
          $[1,0,0]$ & $-0.222222(1)$  \\[0.2ex]
          $[1,1,0]$ & $-0.041164(1)$  \\[0.2ex]
          $[1,1,1]$ & $-0.015484(1)$  \\[0.2ex]
          $[2,0,0]$ & $\phantom{-}0.111111(1)$  \\[0.2ex]
          $[2,2,0]$ & $\phantom{-}0.023521(1)$  \\[0.2ex]
          $[2,2,1]$ & $-0.002620(1)$  \\[0.2ex]
          $[2,2,2]$ & $\phantom{-}0.019055(1)$  \\[0.2ex]
          \hline\hline
        \end{tabular}
        \vskip 0.46cm
        \caption{Coefficients of the static self-energy at one-loop order of PT.\label{fig:statcoefs}}
      \end{center}
      &
      \begin{center}
          \vskip -0.5cm
        \begin{tabular}{cccc}
          \hline\hline\\[-2.0ex]
          action & ${\rm X}$ & $c^{{\rm stat}(1)}_{\rm\scriptscriptstyle X}$ & $b^{{\rm stat}(1)}_{\rm\scriptscriptstyle X}$  \\ \\[-2.0ex]
          \hline\\[-2.0ex]
          HYP1 & A & \phantom{0}0.0029(2) &  \phantom{-}0.0906(2)  \\[0.2ex]
          HYP1 & V & \phantom{-}0.0223(6) &            -0.0212(8)  \\[1.5ex]
          HYP2 & A & \phantom{-}0.0518(2) &  \phantom{-}0.142(1)   \\[0.2ex]
          HYP2 & V & \phantom{-}0.0380(6) &            -0.0462(8)  \\[0.2ex]
          \hline\hline
        \end{tabular}
        \vskip 1.28cm
        \caption{Improvement coefficients of the static-light currents at
          one-loop order of PT.\label{tab:imprcoefs}}
      \end{center}
    \end{tabular}
    \vskip -0.5cm
\end{table}
\end{center}

\vskip -1.0cm
Since \Oa lattice artefacts depend upon all the details of 
the calculation, the information given in the previous paragraphs 
has to be complemented with further technical details, in order to
allow for complete reproducibility. In particular, plots reported 
in Figs.~\ref{fig:cAstatHYP}--\ref{fig:bVstatHYP} refer to a choice 
of the critical quark mass as obtained from the PCAC relation. 
Numerical values used have been taken from  
refs.~\cite{Palombi:2005zd,Palombi:2006pu}. Moreover, the determination 
of $\bAstatone$ requires the one-loop coefficient 
$Z_{\rm\scriptscriptstyle A}^{\rm stat(1)}$ of the renormalisation 
constant of the static axial current as input. The scheme dependence 
of this coefficient has no effect on the continuum limit of $\bAstatone$, 
but it changes its continuum approach. Here, for each single $\theta$-value, 
we used a definition of $Z_{\rm\scriptscriptstyle A}^{\rm stat(1)}$ from 
Eq.~(4.14) of ref.~\cite{Kurth:2000ki}, where the ${\rm O}(a^2)$
$\theta$-dependent corrections at finite lattice spacing are taken as part 
of the definition. This allows for a strong cancellation of the \Oa lattice 
artefacts in $\bAstatone$ and leads to a safe continuum extrapolation. 
Finally, the particular implementation of the axial 
Ward Identity needed for $\cVstatone$, which has been adopted here, refers
to a choice of the lattice topology as ${\cal T}=2$, according to the
notation of ref.~\cite{Palombi:2007dt}.

\begin{center}
  \begin{figure}[!t]
    \begin{center}
      \begin{tabular}{p{0.45\textwidth}p{0.45\textwidth}}
        \epsfig{file=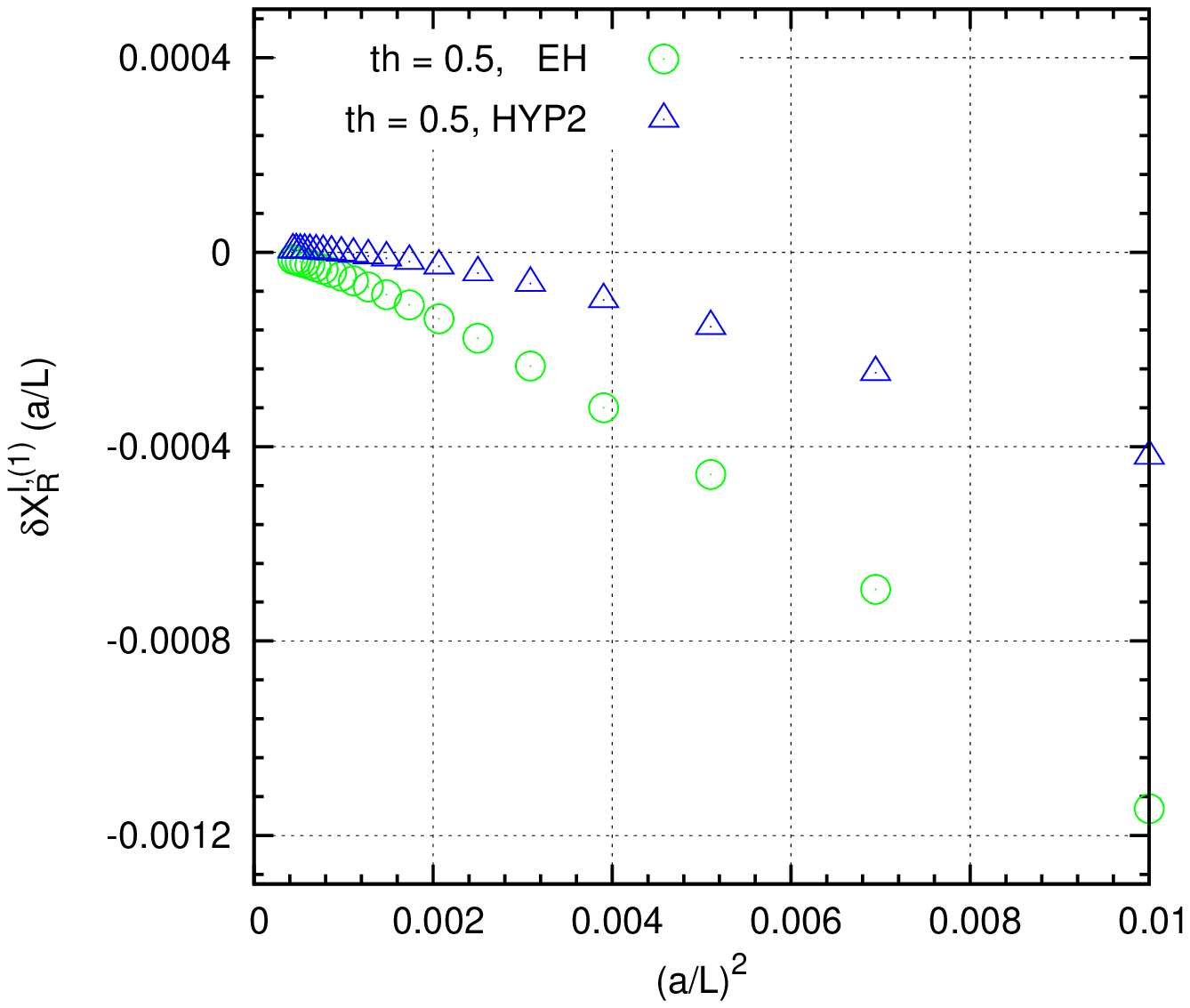,angle=0.0,width=0.45\textwidth}
        &
        \epsfig{file=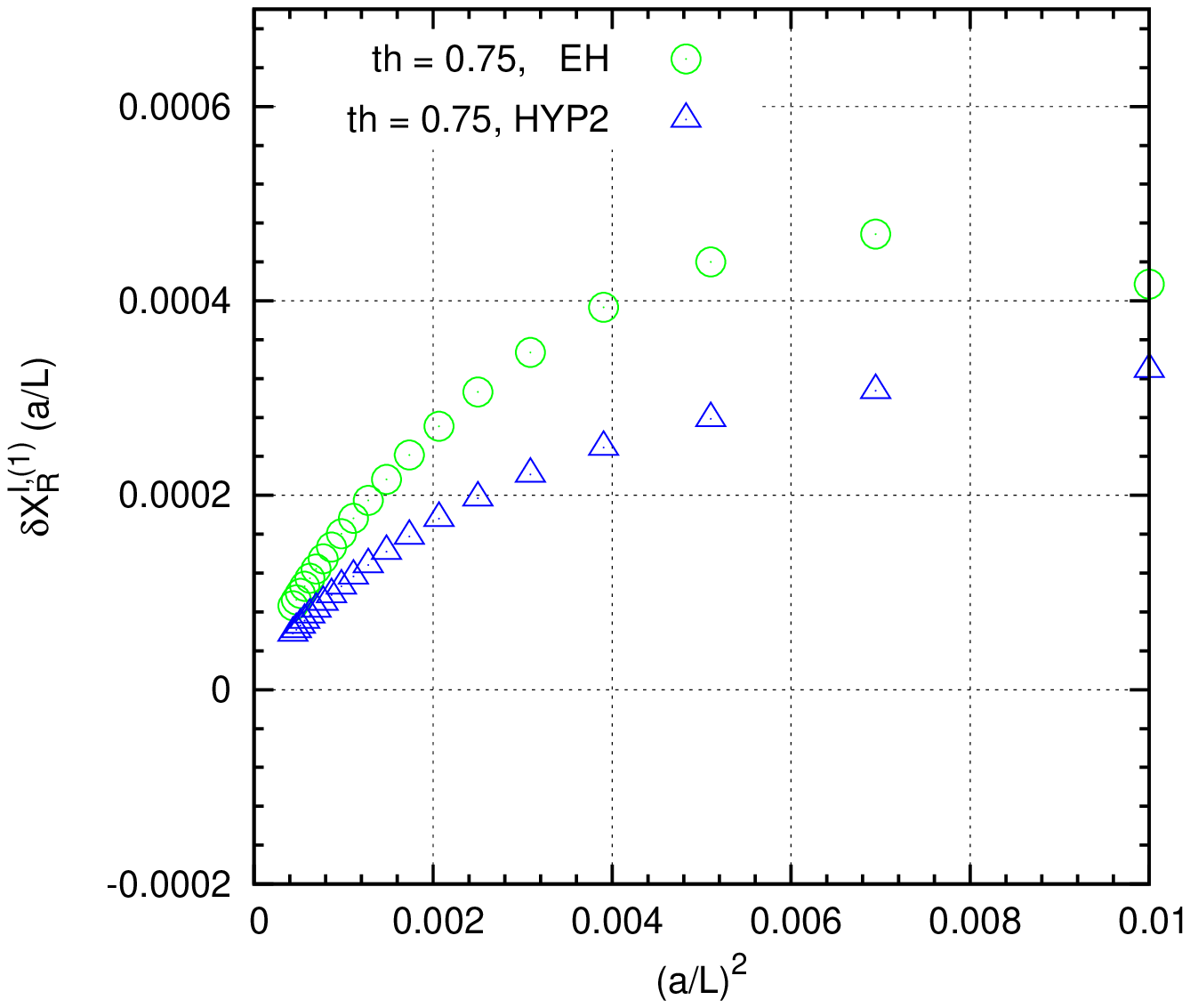,angle=0.0,width=0.45\textwidth}
      \end{tabular}
    \end{center}
    \caption{Comparison of the residual ${\rm O}(a^2)$ lattice artefacts of the 
      improved ratio $X^{\rm\scriptscriptstyle I,(1)}_{\rm\scriptscriptstyle R}$
      at one-loop order of PT in the chiral limit, with $\theta=0.5$ (left) and $\theta=0.75$ (right)
      for the EH (circles) and HYP2 (triangles) actions.
      \label{fig:Otwolat}}
  \end{figure}
\end{center}

Once the improvement coefficients have been determined, one can look at the 
residual ${\rm O}(a^2)$ lattice artefacts of the observables from which those
coefficients are extracted. As an example, we consider the cutoff effects
\begin{equation}
\label{eq:cutoff}
\delta X^{\rm\scriptscriptstyle I}_{\rm\scriptscriptstyle R}(a/L) = \frac{X^{\rm\scriptscriptstyle I}
_{\rm\scriptscriptstyle R}(a/L)}{X^{\rm\scriptscriptstyle I}_{\rm\scriptscriptstyle R}(0)} - 1\ , \qquad
X_{\rm\scriptscriptstyle R}^{\rm\scriptscriptstyle I} = \ZAstat \frac{\,f^{\rm stat}_{\rm\scriptscriptstyle A} + \cAstat
f^{\rm stat}_{\rm\scriptscriptstyle \delta A}\,}
{\sqrt{f_1^{\rm stat}}}\ ,
\end{equation}
of the ratio $X_{\rm\scriptscriptstyle R}^{\rm\scriptscriptstyle I}$ in the chiral limit. 
Here we choose to renormalise the static axial current in the ``lat'' 
scheme, i.e. with $\ZAstat$ having only divergent logarithms without finite parts. Eq.~(\ref{eq:cutoff}) 
can be expandend in PT,
\begin{align}
X_{\rm\scriptscriptstyle R}^{\rm\scriptscriptstyle I} & = X_{\rm\scriptscriptstyle R}^{\rm\scriptscriptstyle I,(0)}
+ g_0^2 X_{\rm\scriptscriptstyle R}^{\rm\scriptscriptstyle I,(1)} + {\rm O}(g_0^4)\ , \\[3.0ex]
\delta X_{\rm\scriptscriptstyle R}^{\rm\scriptscriptstyle I} & = \delta X_{\rm\scriptscriptstyle R}^{\rm\scriptscriptstyle I,(0)}
+ g_0^2 \delta X_{\rm\scriptscriptstyle R}^{\rm\scriptscriptstyle I,(1)} + {\rm O}(g_0^4)\ ,
\end{align}
\begin{align}
\delta X_{\rm\scriptscriptstyle R}^{\rm\scriptscriptstyle I,(0)} & = \frac{X_{\rm\scriptscriptstyle R}^{\rm\scriptscriptstyle I,(0)}\left(a/L\right)}
{X_{\rm\scriptscriptstyle R}^{\rm\scriptscriptstyle I,(0)}\left(0\right)} - 1\ , \\[1.0ex]
\delta X_{\rm\scriptscriptstyle R}^{\rm\scriptscriptstyle I,(1)} & = \left[\frac{X_{\rm\scriptscriptstyle R}^{\rm\scriptscriptstyle I,(1)}\left(a/L\right)}
{X_{\rm\scriptscriptstyle R}^{\rm\scriptscriptstyle I,(0)}\left(a/L\right)} - \frac{X_{\rm\scriptscriptstyle R}^{\rm\scriptscriptstyle I,(1)}\left(0\right)}
{X_{\rm\scriptscriptstyle R}^{\rm\scriptscriptstyle I,(0)}\left(0\right)}\right]\frac{X_{\rm\scriptscriptstyle R}^{\rm\scriptscriptstyle I,(0)}\left(a/L\right)}{X_{\rm\scriptscriptstyle R}^{\rm\scriptscriptstyle I,(0)}\left(0\right)}
\end{align}
\vskip -0.1cm
We note that $\delta X_{\rm\scriptscriptstyle R}^{\rm\scriptscriptstyle I,(0)}$ does not
depend upon the choice of the static regularisation. In order to compare lattice artefacts
corresponding to different actions, one has to consider at least the one-loop contribution 
$\delta X_{\rm\scriptscriptstyle R}^{\rm\scriptscriptstyle I,(1)}$, which is plotted in 
Fig.~\ref{fig:Otwolat} for EH and HYP2 at $\theta \in \{0.5,0.75\}$. 
Remarkably, ${\rm O}(a^2)$ lattice artefacts are less than 0.1\% for both actions. 
Although in principle the smearing of the gauge link could be responsible for an 
enhancement of the cutoff effects, we do not observe any sign of this in our observables. 

\vspace{0.1cm}

It is also important to stress that the perturbative values collected in 
Table~\ref{tab:imprcoefs} need not to be in agreement with the findings
of refs.~\cite{DellaMorte:2005yc,Palombi:2007dt}, where a hybrid technique
has been adopted, which mixes one-loop perturbative inputs obtained from 
the EH or APE discretisations with non-perturbative simulations of  
HYP static fermions. As an example of this, we consider the case of 
$\bVstat$, for which the mixed procedure gives
\begin{alignat}{3}
& b_{\rm\scriptscriptstyle V,HYP1}^{ {\rm stat}} & \ \approx \ \frac{1}{2} -0.014(3)g_0^2 + {\rm O}(g_0^4) \ ,  \\[0.5ex]
& b_{\rm\scriptscriptstyle V,HYP2}^{ {\rm stat}} & \ \approx \ \frac{1}{2} -0.096(8)g_0^2 + {\rm O}(g_0^4) \ .
\end{alignat}
As it can be observed, the exact perturbative coefficients are sensibly
different from the effective ones, which signals the presence
of non-negligible ${\rm O}(g_0^4)$ terms within the latter, quantifiable 
as the differences among the perturbative and the effective values. The hybrid 
procedure is such that the differences absorb ${\rm O}(g_0^4)$ terms
of both actions used in the improvement condition, for the 
limited range of the bare coupling ($6.0 \le \beta \le 6.5$) where this 
has been implemented. A larger discrepancy between the exact improvement 
coefficients of the two discretisations at ${\rm O}(g_0^4)$ induces a larger 
discrepancy between the perturbative coefficients of Table~\ref{tab:imprcoefs} 
and their effective partners. However, we remark that the impact of these ${\rm O}(g_0^4)$
terms is suppressed when $\bVstat$ is multiplied by a reasonably small quark 
mass $am_{\rm q}$. Similar considerations can be done also for the 
other improvement coefficients. 
\vskip 0.5cm

\section{Conclusions}
In this work, we have studied the \Oa improvement of the static
axial and vector currents at one-loop order of perturbation theory 
with hypercubic static and Wilson light quarks. Our methodology for 
the on-shell \Oa improvement is based on the Schr\"odinger 
Functional. The calculation is useful in that it allows to quantify
the amount of the ${\rm O}(g_0^4)$ terms present in the hybrid
perturbative/non-perturbative procedure of 
refs.~\cite{DellaMorte:2005yc,Palombi:2007dt}.

\vspace{0.1cm}

The lack of knowledge of the Feynman rules for the hypercubic static 
propagator in time-momentum representation has prevented such 
calculation over the past years. We simply observed that, 
since the static propagator is a product of temporal hypercubic
links, which are not affected by the boundaries of the Schr\"odinger
Functional, the Feynman rules need not to be necessarily
derived from scratch, but can be obtained by inverse Fourier 
transforming in time the ones obtained in full momentum space, 
the latter being known from the literature
\cite{Hasenfratz:2001tw}.  

\vspace{0.1cm}

Aside the improvement coefficients, we have found an analytical
expression for the static self-energy as a function of the 
smearing parameters at one-loop order of perturbation theory.
This expression has been minimised through an optimal choice of
the smearing parameters. We have found that the perturbative 
minimum is very close to the non-perturbative one, obtained 
by means of an analogous numerical procedure performed at 
non-perturbative values of the bare coupling 
\cite{DellaMorte:2005yc,Sommer:private}. This is interpreted
as a hint of fast perturbative convergence. 
\acknowledgments

We thank M.~Della Morte, A.~Hasenfratz and M.~Papinutto for useful 
discussions. Special thanks go to R.~Hoffmann for sharing 
with us his results of the static self-energy. We are indebted to
R.~Sommer for helping us check the original calculation of $\bAstatone$ 
with Eichten-Hill action and for a careful reading of the draft. F.P. 
acknowledges DESY-Zeuthen for providing hospitality during the initial 
stage of the project. The computing centre of DESY Zeuthen is acknowledged 
for its technical support. This work was supported in part by the EU 
Contract No. MRTN-CT-2006-035482, ``FLAVIAnet''.
 
\appendix

\section{Derivation of the temporal Feynman rules by direct construction}
\label{app:A}

Owing to the irrelevance of the SU(3) projection at one-loop order 
of PT, the HYP smearing procedure can be considered as a three-step 
linear lattice differential operator acting on the gauge fields at 
subsequent levels. The perturbative expansion of 
\hbox{Eqs.~(\ref{hyper1})--(\ref{hyper3})} reads
\begin{align}
\label{hypgauge3}
B_{\mu}^{(3)}(x) & = (1-\alpha_1)q_\mu(x) + 
\frac{\alpha_1}{3}\sum_{\nu\ne\mu}B_{\mu;\nu}^{(2)}(x)\ 
+ \nonumber \\[0.5ex]
& + \frac{\alpha_1}{6}\sum_{\nu\ne\mu}a^2\{\partial^*_\nu\partial_\nu B_{\mu;\nu}^{(2)}(x)
 - \partial_{\nu}^*\partial_\mu B_{\nu;\mu}^{(2)}(x)\}\ + {\rm O}(g_0)\ , \\[1.0ex]
\label{hypgauge2}
B_{\mu;\nu}^{(2)}(x) & = (1-\alpha_2)q_\mu(x) + 
\frac{\alpha_2}{2}\sum_{\rho\ne\mu\nu}B_{\mu;\nu\rho}^{(1)}(x)\ + 
\nonumber \\[0.5ex]
& + \frac{\alpha_1}{4}\sum_{\rho\ne\mu\nu}a^2\{\partial^*_\rho\partial_\rho 
B_{\mu;\nu\rho}^{(1)}(x) - \partial_{\rho}^*\partial_\mu 
B_{\rho;\mu\nu}^{(1)}(x)\}\ + {\rm O}(g_0) \ , \\[0.5ex]
\label{hypgauge1}
B_{\mu;\nu\rho}^{(1)}(x) & = q_\mu(x) + \frac{\alpha_3}{2}\sum_{\eta\ne\mu\nu\rho}
a^2\{\partial^*_\eta\partial_\eta q_{\mu}(x) - \partial_{\eta}^*\partial_\mu q_{\eta}(x)\}
+ {\rm O}(g_0)\ ,
\end{align}
where $\partial_\mu$ and $\partial_\mu^*$ denote respectively the standard 
forward and backward lattice derivatives. 

Since in the SF perturbative calculations are naturally performed in time-momentum 
representation, the above formulae must be Fourier-transformed along the 
spatial directions. Before any explicit calculation, it is worth
first deciding what is relevant to our aims. To be concrete, we are interested
in $\tilde B_0^{(3)}$. Accordingly, the Lorentz sums in \Eq{hypgauge3} are purely 
spatial. The only components of $B_{\mu;\nu}^{(2)}$ to be considered are 
 $B_{0;k}^{(2)}$ and $B_{k;0}^{(2)}$ with $k=1,2,3$. Analogously 
it can be argued of the following steps. If the notation ``$A\leadsto B$'' 
means that $B$ is needed to calculate $A$, then the whole calculation is summarised
by
\begin{alignat}{3}
B_0^{(3)} \ & \leadsto \ B_{0;k}^{(2)}\ , \ B_{k;0}^{(2)} \ , \qquad \qquad
 & k & = 1,2,3 \ ; \\[0.5ex]
B_{0;k}^{(2)} \ & \leadsto \ B_{0;k\ell}^{(1)}\ , \ B_{\ell;0k}^{(1)} \ , 
\qquad \qquad & k & \ne \ell = 1,2,3 \ ;  \\[0.5ex]
B_{k;0}^{(2)} \ & \leadsto \ B_{k;0\ell}^{(1)}\ , \ B_{\ell;0k}^{(1)} \ , 
\qquad \qquad & k & \ne \ell = 1,2,3 \ .
\end{alignat}

In order to obtain $\tilde B_0^{(3)}$ we first consider the Fourier transforms 
of Eqs.~(\ref{hypgauge3})--(\ref{hypgauge1}) separately. Afterwards, 
we insert the expression obtained for the first-level smeared gauge fields 
$\tilde B^{(1)}_{\mu;\nu\rho}$ into $\tilde B^{(2)}_{\mu;\nu}$ and the latter 
into $\tilde B^{(3)}_{0}$. The Fourier transform of the first-level smeared
gauge field is given by
\begin{align}
\tilde B_{0;k\ell}^{(1)}(x_0;\bp) & = \biggl\{1 - \frac{\alpha_3}{2}
\sum_{m\ne k\ell}a^2\hat{p}_m^2\biggr\}\tilde q_0(x_0;\bp) -i\frac{\alpha_3}{2}
\sum_{m\ne k\ell}(a\hat{p}_m)a\partial_0\tilde q_m(x_0;\bp)\ , \\[0.5ex]
\tilde B_{k;0\ell}^{(1)}(x_0;\bp) & = \biggl\{1 - \frac{\alpha_3}{2}
\sum_{m\ne k\ell}a^2\hat{p}_m^2\biggr\}\tilde q_k(x_0;\bp) +\frac{\alpha_3}{2}
\sum_{m\ne k\ell}(a\hat{p}_m)(a\hat{p}_k)\tilde q_m(x_0;\bp)\ , \\[0.5ex]
\tilde B_{\ell;0k}^{(1)}(x_0;\bp) & = \biggl\{1 - \frac{\alpha_3}{2}
\sum_{m\ne k\ell}a^2\hat{p}_m^2\biggr\}\tilde q_\ell(x_0;\bp) +\frac{\alpha_3}{2}
\sum_{m\ne k\ell}(a\hat{p}_m)(a\hat{p}_l)\tilde q_m(x_0;\bp)\ ,
\end{align}
where $\hat p_k \equiv (2/a)\sin(ap_k/2)$ denotes a lattice momentum. In particular, 
it should be remarked that $m\ne k,\ell$, i.e. the sum over 
$m$ contains just one term. Similarly, one finds
\begin{align}
\tilde B_{0;k}^{(2)}(x_0;\bp) & = (1-\alpha_2)\tilde q_0(x_0;\bp) + 
\frac{\alpha_2}{4}\sum_{\ell\ne k}(2-a^2\hat{p}_\ell^2)\tilde 
B_{0;k\ell}^{(1)}(x_0;\bp) \ + \nonumber \\[0.8ex]
& - i\frac{\alpha_2}{4}\sum_{\ell\ne k} (a\hat{p}_\ell)a\partial_0
\tilde B_{\ell;0k}^{(1)}(x_0;\bp)\ , \\[1.5ex]
\tilde B_{k;0}^{(2)}(x_0;\bp) & = (1-\alpha_2)\tilde q_k(x_0;\bp) + 
\frac{\alpha_2}{4}\sum_{\ell\ne k}(2-a^2\hat{p}_\ell^2)\tilde 
B_{k;0\ell}^{(1)}(x_0;\bp) \ + \nonumber \\[0.8ex]
& +\frac{\alpha_2}{4}\sum_{\ell\ne k} (a\hat{p}_\ell)(a\hat{p}_k)
\tilde B_{\ell;0k}^{(1)}(x_0;\bp)\ ,
\end{align}
as for the second-level smeared gauge field. Finally, 
\begin{align}\label{e:B_0_final}
\tilde B_0^{(3)}(x_0;\bp) & = (1-\alpha_1)\tilde q_0(x_0;\bp) + 
\frac{\alpha_1}{6}\sum_{k=1}^3(2-a^2\hat p_k^2)\tilde B^{(2)}_{0;k}(x_0;\bp) \ + \nonumber \\[0.8ex] 
& - i\frac{\alpha_1}{6}\sum_{k=1}^3(a\hat p_k)a\partial_0 \tilde B^{(2)}_{k;0}(x_0;\bp)\ . 
\end{align}
Instead of working out the whole expression alltogether, we observe 
that the final result is expected to be a linear combination of 
$\tilde q_0(x_0;\bp)$, $\tilde q_k(x_0;\bp)$ and 
$\tilde q_k(x_0+a;\bp)$. Single contributions can be considered
separately.

\vspace{0.1cm}

In order to extract the coefficient of $\tilde q_0(x_0,\bp)$, we  
observe that $\tilde B_{k;0\ell}^{(1)}$ and $\tilde B_{\ell;0k}^{(1)}$ have 
no temporal gauge field, which appears only in $\tilde B_{0;k\ell}^{(1)}$. 
Accordingly, $\tilde B^{(2)}_{k;0}$ is independent of $\tilde q_0$. 
Instead, $\tilde B^{(2)}_{0;k}$ has such dependence. The 
coefficient $c_0^{(2)}$ multiplying $\tilde q_0$ in 
$\tilde B^{(2)}_{0;k}$ can be easily isolated, i.e.
\begin{equation}
c_0^{(2)} = 1 - \frac{\alpha_2}{4}(1+\alpha_3)\sum_{\ell\ne k}a^2\hat p_\ell^2 + \frac{\alpha_2\alpha_3}{4}\prod_{\ell\ne k}a^2\hat p_\ell^2 \ . 
\end{equation}
Analogously, the coefficient $c_0^{(3)}$ multiplying 
$\tilde q_0$ in $\tilde B_0^{(3)}$  can be isolated 
upon replacing $\tilde B_{0;k}^{(2)}$ with its explicit 
value. Some algebra leads to 
\begin{equation}\label{e:c_0_3_1}
c_0^{(3)} = (1-\alpha_1) + \frac{\alpha_1}{6}\sum_k (2-a^2\hat p_k^2)c_0^{(2)} = f_{00}(p)\ .
\end{equation}
A similar procedure allows to extract the spatial contributions
to $\tilde B_0^{(3)}(x_0;\bp)$.
\section{Spatial parallel transporter}
\label{app:B}

In this appendix we report the Feynman rules for the spatial HYP link in 
time-momentum representation. We follow sect.~4 and base our derivation 
on the inverse Fourier transform of the Feynman rules in momentum space, 
first obtained in ref.~\cite{Hasenfratz:2001tw}. We start from Eq.~(\ref{eq:decomp}), 
\begin{equation}\label{e:B_k_p}
\tilde B_k^{(3)}(p)=f_{k0}(p)\tilde q_0(p)+\!\sum_{\eta\neq 0,k}\!f_{k\eta}(p)\tilde q_\eta(p)
+f_{kk}(p)\tilde q_k(p)+{\rm O}(g_0)\,.
\end{equation}
Contributions on the right hand side are worked out separately. Some algebra
leads to the expressions
\begin{align}
\label{eq:firstcontr}
\tilde B_k^{(3;1)}(x_0;\bp) & \equiv \frac{1}{L}\sum_{p_0}\re^{ip_0x_0}f_{k0}(p)\tilde q_0(p) = \nonumber \\[1.2ex]
& = -i\frac{\alpha_1}{6}(a\hat p_k)\Omega_{k0}(\bp)a\partial_0^*\tilde q_0(x_0;\bp) \ , \\[3.0ex]
\label{eq:secondcontr}
\tilde B_k^{(3;2)}(x_0;\bp) & \equiv \frac{1}{L}\sum_{p_0}\re^{ip_0x_0}\sum_{\eta\ne 0,k}f_{k\eta}(p)\tilde q_\eta(p) = \nonumber \\[1.2ex]
& = (a\hat p_k)\sum_{\eta\ne 0,k}\left[\Delta_{k\eta}^{\rm (s)}(\bp) + \Delta_{k\eta}^{\rm (t)}(\bp)a^2\partial_0^*\partial_0\right]
\tilde q_\eta(x_0;\bp)\ , \\[3.0ex]
\label{eq:thirdcontr}
\tilde B_k^{(3;3)}(x_0;\bp) & \equiv \frac{1}{L}\sum_{p_0}\re^{ip_0x_0}f_{kk}(p)\tilde q_k(p) = \nonumber \\[1.2ex]
& =\biggl\{1 + \frac{\alpha_1}{6}\Omega_{k0}(\bp)a^2\partial_0^*\partial_0 - \biggr.\nonumber \\[1.2ex]
& - \biggl.\phantom{\int}\sum_{\eta\ne 0,k}(a\hat p_\eta)\left[\Delta_{k\eta}^{\rm (s)}(\bp)
 + \Delta_{k\eta}^{\rm (t)}(\bp)a^2\partial_0^*\partial_0\right]\biggr\}\tilde q_k(x_0;\bp)\ , 
\end{align}
where we have introduced the symbols
\begin{align}
\Delta_{k\eta}^{\rm(s)}(\bp) & = \frac{\alpha_1}{6}(a\hat p_{\eta})\left[ 1 + \alpha_2(1+\alpha_3) -
\frac{\alpha_2}{4}(1+2\alpha_3)(a^2\hat p^2_{\ell\ne k,\eta})\right] \ , \\[1.2ex]
\Delta_{k\eta}^{\rm(t)}(\bp) & = \frac{\alpha_1}{6}(a\hat p_{\eta})\left[\frac{\alpha_2}{4}(1+2\alpha_3) - \frac{
\alpha_2\alpha_3}{4}(a^2\hat p_{\ell\ne k,\eta}^2)\right]\ .
\end{align}
Eqs.(\ref{eq:firstcontr})-(\ref{eq:thirdcontr}) can be written in the compact form
\begin{equation}
\tilde B_k^{(3)}(x_0;\bp) = \sum_{i=0}^{7} V_{k;i}(\bp)\tilde q_{\mu(i)}(x_0+as(i);\bp)\ ,
\end{equation}
where the vertex $V_{k;i}(\bp)$ and the auxiliary indices $\mu$, $s$ are collected in Table~\ref{tab:Feynspatial}.
\begin{table}[!t]
  \begin{center}
    \vbox{\vskip 0.2cm}
    \begin{tabular}{cccc}
      \hline\hline\\[-2.0ex]
      $i$ & $\mu(i)$ & $s(i)$ & $V_{k;i}(\bp)$  \\ \\[-2.0ex]
      \hline\\[-2.0ex]
      0 & 0 & \phantom{-}0 &  $-i\frac{\alpha_1}{6}a\hat p_k\Omega_{k0}(\bp)$ \\[1.5ex]
      1 & 0 & -1 & $\phantom{-}i\frac{\alpha_1}{6}a\hat p_k\Omega_{k0}(\bp)$\\[1.5ex]
      2 & k & \phantom{-}0 &  $1 - \frac{\alpha_1}{3}\Omega_{k0}(\bp)\, + \,\sum_{\eta\neq 0,k}a\hat p_\eta \,[2\Delta_{k\eta}^{\rm (t)}(\bp)-\Delta_{k\eta}^{\rm (s)}(\bp)]$\\[1.5ex]
      3 & k & \phantom{-}1 & $\frac{\alpha_1}{6}\Omega_{k0}(\bp) - \sum_{\eta\neq 0,k}a\hat p_\eta\Delta_{k\eta}^{\rm (t)}(\bp)$\\[1.5ex]
      4 & k & -1 &  $\frac{\alpha_1}{6}\Omega_{k0}(\bp) - \sum_{\eta\neq 0,k}a\hat p_\eta\Delta_{k\eta}^{\rm (t)}(\bp)$\\[1.5ex]
      5 & $\eta;0k$ & \phantom{-}0 &  $a\hat p_k[\Delta_{k\eta}^{(s)}(\bp) -2\Delta_{k\eta}^{\rm(t)}(\bp)]$\\[1.5ex]
      6 & $\eta;0k$ & \phantom{-}1 &  $a\hat p_k \Delta_{k\eta}^{\rm(t)}(\bp)$\\[1.5ex]
      7 & $\eta;0k$ & -1 &   $a\hat p_k \Delta_{k\eta}^{\rm(t)}(\bp)$\\[1.5ex]
      \hline\hline
    \end{tabular}
    \caption{Feynman rules of the spatial HYP link in time-momentum representation. The symbol $\eta;0k$ 
      denotes a sum over $\eta\ne 0,k$.\label{tab:Feynspatial}}
  \end{center}
\end{table}

\section{Derivation of Eq.~(\protect\ref{eoneres})}
\label{app:C}

To conclude, we detail the calculation of the static self-energy and provide
a derivation of \Eq{eoneres}. We consider the 
boundary-to-boundary correlator $f_1^{\rm stat}$, defined in 
ref.~\cite{Kurth:2000ki} via
\begin{equation}
f_1^{\rm stat} = -\frac{a^{12}}{2L^6}\sum_{\bu\bv\by\bz}\langle \overline{\zeta}'(\bu)\gamma_5
\zeta'_{\rm h}(\bv)\overline{\zeta}_{\rm h}(\by)\gamma_5
\zeta(\bz)\rangle \ . 
\end{equation}
Expanding $f_1^{\rm stat}$ at one-loop order of PT allows to write the  
coefficient $e^{(1)}$, see \Eq{eonedef}, in terms of the Feynman diagrams of Fig.~\ref{fig:Feynman}, i.e.
\begin{align}
\label{ecoef}
e^{(1)} & = -\lim_{a/L\to0}\frac{4}{3}\frac{a}{L^4}\sum_\bp a^2 \sum_{u_0=a}^{T}\sum_{v_0=a}^{u_0} b(u_0,v_0) \sum_{i,j=0}^6 \delta_{\mu(i)\mu(j)} V_{0;i}(\bp)V_{0;j}(-\bp)\ \times \nonumber \\[2.0ex]
& \times d_{\mu(i)\mu(j)}(u_0-a + as(i),v_0-a+as(j);\bp)\ = \nonumber \\[2.0ex]
& \equiv -\lim_{a/L\to0}\frac{4}{3}\frac{1}{L^3}\sum_\bp a^2 \sum_{u_0=a}^{T}\sum_{v_0=a}^{u_0} b(u_0,v_0){\cal V}(u_0,v_0,\bp) \ , 
\end{align}
where $d_{\mu\nu}(x_0,y_0,\bp)$ denotes the gluon propagator in time-momentum representation, cf.
ref.~\cite{Luscher:1996vw} for a definition. The weight-coefficient $b(u_0,v_0)$ is given by
\begin{equation}
b(u_0,v_0) = \left\{  \begin{array}{cl}
    1/2 & \quad u_0 = v_0 \ ,\\[1.2ex]
    1   & \quad {\rm otherwise} \ , \end{array} \right .
\end{equation}
and the interaction blob 
\begin{equation}
{\cal V}(u_0,v_0,\bp) = \sum_{i,j=0}^6 \delta_{\mu(i)\mu(j)} V_{0;i}(\bp)V_{0;j}(-\bp)d_{\mu(i)\mu(j)}(u_0-a + as(i),v_0-a+as(j);\bp)
\end{equation}
denotes a HYP gluon propagating on the lattice from time $u_0$ to time $v_0$ with spatial momentum $\bp$. 
The above expression can be simplified by using spatial rotational invariance, i.e. $d_{11} = d_{22} = d_{33}$
and $d_{ij} = 0$ if $i\ne j$. Accordingly, the vertex reads
\begin{align}
\label{vertex}
{\cal V}(u_0,v_0,\bp) & = |V_{0;0}(\bp)|^2d_{00}(u_0-a,v_0-a;\bp)\ + \nonumber \\[2.0ex]
& + \left[|V_{0;1}(\bp)|^2 + |V_{0;2}(\bp)|^2 + |V_{0;3}(\bp)|^2\right]\ \times \nonumber \\[2.0ex]
& \times \left[ d_{kk}(u_0-a,v_0-a;\bp) + d_{kk}(u_0,v_0;\bp) \right. \ - \nonumber \\[2.0ex] 
& - \left. d_{kk}(u_0,v_0-a;\bp) - d_{kk}(u_0-a,v_0;\bp) \right]\ . 
\end{align}
From \Eq{vertex}, we conclude that the HYP vertex $V_{0;j}(\bp)$ enters 
the coefficient $e^{(1)}$ only in the rotationally symmetric combinations
\begin{align}
h^{\rm (t)}(\bp) & = |V_{0;0}(\bp)|^2 \ , \nonumber \\[1.5ex]
h^{\rm (s)}(\bp) & = |V_{0;1}(\bp)|^2 + |V_{0;2}(\bp)|^2 + |V_{0;3}(\bp)|^2\ . 
\end{align}
These quantities are multivariate polynomials of 
the HYP smearing parameters, i.e.  
\begin{equation}
\label{hts}
h^{\rm(t,s)}(\bp) = \sum_{k_1k_2k_3=0}^2 w^{\rm(t,s)}_{k_1k_2k_3}(\bp)\alpha_1^{k_1}\alpha_2^{k_2}\alpha_3^{k_3}\ , \qquad w^{\rm(t,s)}_{k_1k_2k_3} = \frac{1}{k_1!k_2!k_3!}\frac{\partial^{k_1+k_2+k_3}h^{\rm(t,s)}}{\partial\alpha_1^{k_1}\partial\alpha_2^{k_2}\partial\alpha_3^{k_3}} \ .
\end{equation}
The coefficients $w^{\rm(t,s)}_{k_1k_2k_3}$ have been written 
according to the generic Taylor expansion in several variables. They
can be algebraically evaluated; the non-vanishing ones are reported 
in sect. C.1 in units of the lattice spacing. Spatial rotational invariance 
is evident. 

Upon inserting \Eq{hts} into \Eq{vertex} and \Eq{vertex} into \Eq{ecoef}, 
the final result represented by \Eq{eoneres} is obtained with coefficients 
\begin{align}
& e^{(1)}_{k_1k_2k_3} = \lim_{a/L\to 0} \frac{4}{3}\frac{a}{L^4}\sum_\bp a^2 \sum_{u_0=a}^{T}\sum_{v_0=a}^{u_0} b(u_0,v_0) \ \times \nonumber \\[1.5ex]
& \times \left\{ w^{\rm(t)}_{k_1k_2k_3}(\bp)d_{00}(u_0-a,v_0-a;\bp) \ + \right\} \nonumber \\[1.5ex]
& +  w^{\rm(s)}_{k_1k_2k_3}(\bp)\left[d_{kk}(u_0-a,v_0-a;\bp) + d_{kk}(u_0,v_0;\bp) \ - \right. \nonumber \\[2.0ex]
& \left.\left. -  d_{kk}(u_0,v_0-a;\bp) - d_{kk}(u_0-a,v_0;\bp)\right]\right\} \ . 
\end{align}

\newpage

\subsection{Coefficients $w_{\rm t,s}^{k_1k_2k_3}$}

\begin{align}
w^{\rm(t)}_{{000}} & = 1\ ,  \\[1.0ex]
w^{\rm(t)}_{{100}} & = -\frac{1}{3}\,({{\hat p_1}}^{2} + {{\hat p_2}}^{2} + {{\hat p_3}}^{2})\ ,  \\[1.0ex]
w^{\rm(t)}_{{110}} & = -\frac{1}{3}\,({{\hat p_1}}^{2}+{{\hat p_2}}^{2} + {{\hat p_3}}^{2}) + \frac{1}{6}\,({{\hat p_1}}^{2}{{\hat p_2}}^{2} + {{\hat p_1}}^{2}{{\hat p_3}}^{2}+{{\hat p_2}}^{2}{{\hat p_3}}^{2})\ ,  \\[1.0ex]
w^{\rm(t)}_{{111}} & = -\frac{1}{3}\,({{\hat p_1}}^{2} + {{\hat p_2}}^{2} + {{\hat p_3}}^{2})+\frac{1}{3}\,({{\hat p_1}}^{2}{{\hat p_2}}^{2} + {{\hat p_1}}^{2}{{\hat p_3}}^{2} + {{\hat p_2}}^{2}{{\hat p_3}}^{2}) -\frac{1}{4}\,{{\hat p_1}}^
{2}{{\hat p_2}}^{2}{{\hat p_3}}^{2}\ ,  \\[1.0ex]
w^{\rm(t)}_{{200}} & = \frac{1}{36}\,({{\hat p_1}}^{4} + {{\hat p_2}}^{4} + {{\hat p_3}}^{4}) + \frac{1}{18}\,({{\hat p_1}}^{2}{{\hat p_2}}^{2}+ {{\hat p_1}}^{2}{{\hat p_3}}^{2}+{{\hat p_2}}^{2}{{\hat p_3}}^{2})\ ,  \\[1.0ex]
w^{\rm(t)}_{{210}} & = \frac{1}{18}({{\hat p_1}}^{4}+ {{\hat p_2}}^{4} + {{\hat p_3}}^{4}) + 
\frac{1}{9}\,({{\hat p_1}}^{2}{{\hat p_2}}^{2}+ {{\hat p_1}}^{2}{{\hat p_3}}^{2}+ {{\hat p_2}}^{2}{{\hat p_3}}^{2})\ - \nonumber \\[1.2ex] 
& -\frac{1}{36}\,({{\hat p_1}}^{4}{{\hat p_2}}^{2} + {{\hat p_1}}^{4}{{\hat p_3}}^{2} + {{\hat p_2}}^{4}{{\hat p_1}}^{2} + {{\hat p_2}}^{4}{{\hat p_3}}^{2} + {{\hat p_3}}^{4}{{\hat p_1}}^{2} + {{\hat p_3}}^{4}{{\hat p_2}}^{2}) -\frac{1}{12}\,{{\hat p_1}}^{2}{{\hat p_2}}^{2}{{\hat p_3}}^{2}\ ,  \\[1.0ex]
w^{\rm(t)}_{{211}} & = \frac{1}{18}\,({{\hat p_1}}^{4} + {{\hat p_2}}^{4} + {{\hat p_3}}^{4}) + \frac{1}{9}\,({{\hat p_1}}^{2}{{\hat p_2}}^{2}
 + {{\hat p_1}}^{2}{{\hat p_3}}^{2} + {{\hat p_2}}^{2}{{\hat p_3}}^{2})\ - \nonumber   \\[1.2ex]
& - \frac{1}{18}\,({{\hat p_1}}^{4}{{\hat p_2}}^{2} + {{\hat p_1}}^{4}{{\hat p_3}}^{2} + {{\hat p_2}}^{4}{{\hat p_1}}^{2} + {{\hat p_2}}^{4}{{\hat p_3}}^{2} + {{\hat p_3}}^{4}{{\hat p_1}}^{2} + {{\hat p_3}}^{4}{{\hat p_2}}^{2}) - \frac{1}{6}\,{{\hat p_1}}^{2}{{\hat p_2}}^{2}{{\hat p_3}}^{2}\ + \nonumber  \\[1.2ex] 
& +\frac{1}{24}\,({{\hat p_1}}^{4}{{\hat p_2}}^{2}{{\hat p_3}}^{2} + {{\hat p_2}}^{4}{{\hat p_1}}^{2}{{\hat p_3}}^{2} + {{\hat p_3}}^{4}{{\hat p_1}}^{2}{{\hat p_2}}^{2}) \ , \\[1.0ex]
w^{\rm(t)}_{{220}} & = \frac{1}{36}\,({{\hat p_1}}^{4} + {{\hat p_2}}^{4} + {{\hat p_3}}^{4}) + \frac{1}{18}(\,{{\hat p_1}}^{2}{{\hat p_2}}^{2}+ {{\hat p_1}}^{2}{{\hat p_3}}^{2} + {{\hat p_2}}^{2}{{\hat p_3}}^{2}) - \frac{1}{12}\,{{\hat p_1}}^{2}{{\hat p_2}}^{2}{{\hat p_3}}^{2}\ - \nonumber \\[1.2ex]
& - \frac{1}{36}\,({{\hat p_1}}^{4}{{\hat p_2}}^{2} + {{\hat p_1}}^{4}{{\hat p_3}}^{2} + {{\hat p_2}}^{4}{{\hat p_1}}^{2} + \,{{\hat p_2}}^{4}{{\hat p_3}}^{2} + {{\hat p_3}}^{4}{{\hat p_1}}^{2} + {{\hat p_3}}^{4}{{\hat p_2}}^{2})\ + \nonumber \\[1.2ex]
& + \frac {1}{72}\,({{\hat p_1}}^{4}{{\hat p_2}}^{2}{{\hat p_3}}^{2} + {{\hat p_2}}^{4}{{\hat p_1}}^{2}{{\hat p_3}}^{2} + {{\hat p_3}}^{4}{{\hat p_1}}^{2}{{\hat p_2}}^{2}) + \frac {1}{144}\,({{\hat p_1}}^{4}{{\hat p_2}}^{4} + {{\hat p_1}}^{4}{{\hat p_3}}^{4} + {{\hat p_2}}^{4}{{\hat p_3}}^{4}) \ , \\[1.0ex]
w^{\rm(t)}_{221} & = \frac{1}{18}\,({\hat p_1}^{4} + {\hat p_2}^{4} + {\hat p_3}^{4}) + \frac{1}{9}\,({\hat p_1}^{2}{\hat p_2}^{2} + {\hat p_1}^{2}{\hat p_3}^{2} + {\hat p_2}^{2}{\hat p_3}^{2})\ - \nonumber \\[1.2ex]
& -\frac{1}{12}\,({\hat p_1}^{4}{\hat p_2}^{2} + {\hat p_1}^{4}{\hat p_3}^{2} + {\hat p_2}^{4}{\hat p_1}^{2} + {\hat p_2}^{4}{\hat p_3}^{2} + {\hat p_3}^{4}{\hat p_1}^{2} + {\hat p_3}^{4}{\hat p_2}^{2}) - \frac{1}{4}\,{\hat p_1}^{2}{\hat p_2}^{2}{\hat p_3}^{2}\ + \nonumber \\[1.2ex]
& + \frac{1}{36}\,({\hat p_1}^{4}{\hat p_2}^{4} + {\hat p_1}^{4}{\hat p_3}^{4} + {\hat p_2}^{4}{\hat p_3}^{4}) + \frac {7}{72}\,({\hat p_1}^{4}{\hat p_2}^{2}{\hat p_3}^{2} + {\hat p_2}^{4}{\hat p_1}^{2}{\hat p_3}^{2} + {\hat p_3}^{4}{\hat p_1}^{2}{\hat p_2}^{2})\ - \nonumber \\[1.2ex]
& - \frac{1}{48}\,({\hat p_1}^{4}{\hat p_2}^{4}{\hat p_3}^{2} + {\hat p_1}^{4}{\hat p_3}^{4}{\hat p_2}^{2} + {\hat p_2}^{4}{\hat p_3}^{4}{\hat p_1}^{2})\ , \\[1.0ex]
w^{\rm(t)}_{222} & = \frac{1}{36}\,( {\hat p_1}^{4} + {\hat p_2}^{4} + {\hat p_3}^{4}) + \frac{1}{18}\,({\hat p_1}^{2}{\hat p_2}^{2} + {\hat p_1}^{2}{\hat p_3}^{2} + {\hat p_2}^{2}{\hat p_3}^{2})\ -  \nonumber \\[1.2ex] 
& -\frac{1}{6}\,{\hat p_1}^{2}{\hat p_2}^{2}{\hat p_3}^{2}-\frac{1}{18}\,({\hat p_1}^{4}{\hat p_2}^{2} + {\hat p_1}^{4}{\hat p_3}^{2} + {\hat p_2}^{4}{\hat p_1}^{2} + {\hat p_2}^{4}{\hat p_3}^{2} + {\hat p_3}^{4}{\hat p_1}^{2} + {\hat p_3}^{4}{\hat p_2}^{2}) \ + \nonumber \\[1.2ex]
& + \frac{1}{36}\,({\hat p_1}^{4}{\hat p_2}^{4} + {\hat p_1}^{4}{\hat p_3}^{4} + {\hat p_2}^{4}{\hat p_3}^{4}) + \frac {7}{72}\,({\hat p_1}^{4}{\hat p_2}^{2}{\hat p_3}^{2} + {\hat p_2}^{4}{\hat p_1}^{2}{\hat p_3}^{2} + {\hat p_3}^{4}{\hat p_1}^{2}{\hat p_2}^{2}) \ - \nonumber \\[1.2ex]
& - \frac{1}{24}\,({\hat p_1}^{4}{\hat p_2}^{4}{\hat p_3}^{2} + {\hat p_1}^{4}{\hat p_3}^{4}{\hat p_2}^{2} + {\hat p_2}^{4}{\hat p_3}^{4}{\hat p_1}^{2}) \ +  \frac {1}{64}\,{\hat p_1}^{4}{\hat p_2}^{4}{\hat p_3}^{4}\ . 
\end{align}

\begin{align}
w^{\rm(s)}_{200} & = \frac{1}{18}\,({\hat p_1}^{2} + {\hat p_2}^{2} + {\hat p_3}^{2})\ ,  \\[1.0ex]
w^{\rm(s)}_{210} & = \frac{1}{9}\,({\hat p_1}^{2} + {\hat p_2}^{2} + {\hat p_3}^{2}) - \frac{1}{18}\,({\hat p_1}^{2}{\hat p_2}^{2} + {\hat p_1}^{2}{\hat p_3}^{2} + {\hat p_2}^{2}{\hat p_3}^{2})\ ,  \\[1.0ex]
w^{\rm(s)}_{211} & = \frac{1}{9}\,({\hat p_1}^{2} + {\hat p_2}^{2} + {\hat p_3}^{2}) -\frac{1}{9}(\,{\hat p_1}^{2}{\hat p_2}^{2} + {\hat p_1}^{2}{\hat p_3}^{2} + {\hat p_2}^{2}{\hat p_3}^{2}) + \frac{1}{12}\,{\hat p_1}^{2}{\hat p_2}^{2}{\hat p_3}^{2}\ ,  \\[1.0ex]
w^{\rm(s)}_{220} & = \frac{1}{9}\,({\hat p_1}^{2} + {\hat p_2}^{2} + {\hat p_3}^{2}) - \frac{1}{9}\,({\hat p_1}^{2}{\hat p_2}^{2} + {\hat p_1}^{2}{\hat p_3}^{2} + {\hat p_2}^{2}{\hat p_3}^{2})\ + \nonumber \\[1.2ex]
& +  \frac{1}{144}\,( {\hat p_1}^{4}{\hat p_2}^{2} + {\hat p_1}^{4}{\hat p_3}^{2} + {\hat p_2}^{4}{\hat p_1}^{2} + {\hat p_2}^{4}{\hat p_3}^{2} + {\hat p_3}^{4}{\hat p_1}^{2} + {\hat p_3}^{4}{\hat p_2}^{2}) +\frac{1}{24}\,{\hat p_1}^{2}{\hat p_2}^{2}{\hat p_3}^{2}\ ,  \\[1.0ex]
w^{\rm(s)}_{221} & = \frac{2}{9}\,({\hat p_1}^{2} + {\hat p_2}^{2} + {\hat p_3}^{2}) - \frac{1}{3}\,({\hat p_1}^{2}{\hat p_2}^{2} + {\hat p_1}^{2}{\hat p_3}^{2} + {\hat p_2}^{2}{\hat p_3}^{2})\ + \nonumber \\[1.2ex]
& + \frac{1}{36}\,( {\hat p_1}^{4}{\hat p_2}^{2} + {\hat p_1}^{4}{\hat p_3}^{2} + {\hat p_2}^{4}{\hat p_1}^{2} + {\hat p_2}^{4}{\hat p_3}^{2} + {\hat p_3}^{4}{\hat p_1}^{2} + {\hat p_3}^{4}{\hat p_2}^{2}) +\frac{1}{3}\,{\hat p_1}^{2}{\hat p_2}^{2}{\hat p_3}^{2}\ - \nonumber \\[1.2ex]
& - \frac{1}{36}\,( {\hat p_1}^{4}{\hat p_2}^{2}{\hat p_3}^{2} + {\hat p_2}^{4}{\hat p_1}^{2}{\hat p_3}^{2} + {\hat p_3}^{4}{\hat p_1}^{2}{\hat p_2}^{2} ) \ ,  \\[1.0ex]
w^{\rm(s)}_{222} & = \frac{2}{9}\,({\hat p_1}^{2} + {\hat p_2}^{2} + {\hat p_3}^{2}) - \frac{4}{9}\,({\hat p_1}^{2}{\hat p_2}^{2} + {\hat p_1}^{2}{\hat p_3}^{2} + {\hat p_2}^{2}{\hat p_3}^{2})\ +  \nonumber \\[1.2ex]
& + \frac{1}{18}\,( {\hat p_1}^{4}{\hat p_2}^{2} + {\hat p_1}^{4}{\hat p_3}^{2} + {\hat p_2}^{4}{\hat p_1}^{2}+ {\hat p_2}^{4}{\hat p_3}^{2} + {\hat p_3}^{4}{\hat p_1}^{2} + {\hat p_3}^{4}{\hat p_2}^{2}) +\frac{2}{3}\,{\hat p_1}^{2}{\hat p_2}^{2}{\hat p_3}^{2} \ +  \nonumber \\[1.2ex]  
& -\frac{1}{9}\,( {\hat p_1}^{4}{\hat p_2}^{2}{\hat p_3}^{2} + {\hat p_2}^{4}{\hat p_1}^{2}{\hat p_3}^{2} + {\hat p_3}^{4}{\hat p_1}^{2}{\hat p_2}^{2}) + \frac{1}{72}\,({\hat p_1}^{4}{\hat p_2}^{4}{\hat p_3}^{2} + {\hat p_1}^{4}{\hat p_3}^{4}{\hat p_2}^{2} + {\hat p_2}^{4}{\hat p_3}^{4}{\hat p_1}^{2})\ . 
\end{align}

\vfill\eject

\end{document}